%% file: Arxiv_Main_v2.tex
\documentclass[twoside, twocolumn, letterpaper, 10pt]{article}

\DeclareMathSizes{10}{10.5}{6.5}{6.5}

\usepackage{anysize}
\usepackage{fancyhdr}

\marginsize{1.75cm}{1.75cm}{0.5cm}{0.5cm}

\input{Packages}

\input{Commands}
\input{Acronyms}

\titleformat*{\section}{\large\bfseries}
\titleformat*{\subsection}{\bfseries}
\begin{document}

	\title{
	\vspace{-1.5cm}\rule{\linewidth}{4pt}\vspace{0.3cm} \Large \textbf{
    Learning with SASQuaTCh: a Novel Variational Quantum Transformer Architecture with Kernel-Based Self-Attention
	}\\ \rule{\linewidth}{1.5pt}}
	\author{Ethan N. Evans\thanks{Corresponding author. Email: \href{ethan.n.evans.civ@us.navy.mil}{ethan.n.evans.civ@us.navy.mil}},\;
    Matthew Cook, Zachary P. Bradshaw, and Margarite L. LaBorde\\
    \vspace{-0.1cm}
	\small{Naval Surface Warfare Center, Panama City Division} \\ \vspace{-0.2cm}
	}
        \date{\vspace{-0.6cm}}
	\maketitle

\vspace{10em}

\begin{abstract}

The recent exploding growth in size of state-of-the-art machine learning models highlights a well-known issue where exponential parameter growth, which has grown to trillions as in the case of the Generative Pre-trained Transformer (GPT), leads to training time and memory requirements which limit their advancement in the near term. The predominant models use the so-called transformer network and have a large field of applicability, including predicting text and images, classification, and even predicting solutions to the dynamics of physical systems. Here we present a variational quantum circuit architecture named \gls{SASQuaTCh}, which builds networks of qubits that perform analogous operations of the transformer network, namely the keystone self-attention operation, and leads to an exponential improvement in parameter complexity and run-time complexity over its classical counterpart. Our approach leverages recent insights from kernel-based operator learning in the context of predicting spatiotemporal systems to represent deep layers of a vision transformer network using simple gate operations and a set of multi-dimensional quantum Fourier transforms. To validate our approach, we consider image classification tasks in simulation and with hardware, where with only 9 qubits and a handful of parameters we are able to simultaneously embed and classify a grayscale image of handwritten digits with high accuracy.

\end{abstract}


\section*{Introduction and Related Work}\label{sec:intro}

The field of \gls{AIML} is becoming increasingly dominated by a 2017 emergent neural network architecture called the transformer network \cite{vaswani2017attention}. The advent and widespread success of this approach, as popularized by ChatGPT \cite{achiam2023gpt}, has lead to revolutionary effects on how people engage with technology, and a race for ever more powerful \glspl{LLM} that reaches milestones on a monthly basis. Nevertheless, \glspl{LLM} are plagued by an exponential growth rate in the number of trainable parameters, which have ballooned to the trillions, with training time and run-time complexity following similar trends. Thus several attempts seek to replicate the performance of trillion-scale networks with three orders-of-magnitude fewer parameters, leading to reduced models with only billions of parameters (e.g. Mistral \cite{jiang2023mistral}, Alpaca \cite{taori2023stanford}, etc.), however the fundamental issue of parameter scalability remains. These exponential trends were recently captured empirically by so-called neural scaling laws \cite{kaplan2020scaling}, and a recent understanding of skill generalization emergence \cite{arora2023theory} suggests a necessity of large models in the classical regime.

\parag Since its advent, applications have been found for transformers outside of NLP, including image recognition \cite{dosovitskiy2021image}, speech recognition \cite{radford2023robust}, and predicting dynamical systems \cite{geneva2022transformers}. The success of the transformer architecture has been largely attributed to its use of a multi-head attention mechanism performing scaled dot-product attention in each unit, and this involves learning three weight matrices, known as the query, key, and value weights for each attention head. This mechanism, along with a positional encoding, allows the model to understand the context of a component of the input in relation to the remaining components.

\parag Quantum computing has arisen as an alternate computing paradigm in which the properties of quantum theory are leveraged to potentially outperform classical devices on certain problems, including prime factorization \cite{shor1997polynomial} and unstructured search problems \cite{grover1996fast}. While such algorithms will have profound consequences in areas such as cryptography \cite{bennett2014cryptography} when a machine capable of running them becomes available, that time has not yet come. In recent years, focus has been on \gls{NISQ} devices \cite{preskill2018quantum} and finding potential uses for these devices today. A promising contender is the rising field of \gls{QML} \cite{schuld2021machine}, which intends to combine the success of machine learning algorithms with the framework of quantum computing to produce better machine learning models by leveraging properties such as superposition and entanglement. The use of \gls{QML} is motivated by universal approximation properties \cite{goto2021universal,jager2023universal}, and despite many challenges \cite{cerezo2022challenges,thanasilp2023subtleties} several approaches boast quantum advantage over classical methods \cite{liu2021rigorous,saggio2021experimental,lloyd2013quantum,aimeur2013quantum}.

\parag In this work, we implement a quantum vision transformer inspired by the \gls{FNO} \cite{li2020fourier}, the \gls{AFNO} \cite{guibas2021adaptive}, and the FourCastNet \cite{pathak2022fourcastnet}. These models make use of the observation that the scaled dot product self-attention mechanism can be represented by a convolution against a stationary kernel. A similar approach was developed in \cite{leethorp2022fnet}, wherein despite their desire to replace the self-attention mechanism with Fourier transforms, the authors ultimately produce a nearly identical architecture with self-attention by means of kernel convolution.

\parag To construct our model, we make use of the \gls{QFT}, which can be implemented efficiently on a quantum computer. The \gls{QFT} performs a version of channel mixing, and enables one to easily perform token mixing of a sequence of tokens via a variational quantum circuit acting as a kernel in the Fourier domain. We call our overall circuit \gls{SASQuaTCh}. While we use image classification to evaluate our approach, this approach is very general in its applicability to sequence data of any form (e.g. solutions to dynamical systems, natural language processing, and time series data).

\parag The \gls{QFT} manipulates the $2^n$ complex amplitudes in an $n$-qubit state with $O(n^2)$ Hadamard gates and controlled phase shift gates, although an efficient approximation can be achieved with $O(n\log(n))$ gates, assuming that controlled phases are native to the quantum computing architecture \cite{hales2000improved}. Meanwhile, the fast Fourier transform on a vector with $2^n$ coefficients takes $O(n2^n)$ operations, so that an apparent exponential speedup has been achieved. However, the \gls{QFT} is not without difficulties; the Fourier coefficients are hidden by measurement collapse and thus can only be estimated via repeated trials. Furthermore, performing the \gls{QFT} requires a nontrivial embedding of the classical data into the amplitudes of a quantum state for which a general efficient algorithm is not presently known, although there are some classes of states for which an efficient preparation is known, e.g. uniform superpositions \cite{levin2023optimized,shukla2024}.

\parag It should be noted that the idea of implementing the self-attention mechanism on a quantum device was taken up in \cite{li2022quantum} with the introduction of the \gls{QSANN}. The quantum self-attention mechanism employed in \gls{QSANN} makes use of three separate circuit ansatze; namely, the query, key, and value parts, which use angle embeddings to embed the prior layer and are subsequently measured. Notably, these measurement outputs are used to \textit{classically} compute the self-attention mechanism. Thus, any quantum advantage from \gls{QSANN} must come from the calculation of the query, key, and value components. Furthermore, the repeated embed-evolve-measure scheme leads to substantial information losses due to repeated projective measurements. Similarly, in \cite{zhao2024qksan}, the authors propose a self-attention mechanism by performing measurement-conditioned variational operations, however it is also unclear how layering can be performed without succumbing to information losses at each layer. These are in contrast to the approach presented here, in which the entirety of the self-attention mechanism is performed on a quantum device, and thus layering by extension is achieved by simply adding kernel layers to the circuit. 

\parag Additional efforts to implement a GPT-style algorithm were proposed while this manuscript was in preparation \cite{Liao2024GPT}; however, the implementation in that work requires extensive gate operations and does not utilize the kernel-based self-attention framework as applied here. Furthermore, we are able to provide proof-of-concept hardware implementations of our work which are not feasible for other, similar works at this time. We also note the recent work \cite{xuemeasurement} which implements multi-head attention in simulation; however, the approach requires the use of quantum memory, which is not presently feasible.

\section*{Results}
\vspace{-1em}
\noindent \textbf{Self-Attention and the Transformer Neural Network}

\noindent The transformer network architecture has quickly become a staple in advanced machine learning. The primary mechanism is known as self-attention, or scaled dot product attention, and the primary data type that transformers are applied to is sequence data. Let $\{x_s \in \Rb^d\}_{s=1}^N$ be a sequence of $N$ tokens representing the input data and let $x \in \Rb^{N\times d}$ represent the entire sequence in tensor notation. As originally introduced in \cite{vaswani2017attention}, the self-attention mechanism is given by
\begin{subequations}\label{eq:attention}
\begin{align}
        a_{s,j} &= \frac{\exp\big(x_s^\T W_q^\T W_k x_j\big)}{\sum_{l=1}^S \exp\big(x_s^\T W_q^\T W_k x_l\big)} \label{eq:normed_exp_attention} \\
    y_s &= \text{Att}(x_s) := \sum_{j=1}^S a_{s,j} W_v x_j,
\end{align}
\end{subequations}
where the output $y_s$ belongs to a sequence $\{y_s \in \Rb^d \}_{s=1}^N$ with similar tensor notation $y \in \Rb^{N\times d}$, and the weights $W_q, W_k, W_v \in \Rb^{d\times d}$ are the trainable query, key, and value matrices, respectively.

The normalized exponential in \cref{eq:normed_exp_attention} applies an asymmetric weighting of the input sequence, and can be represented as a scaled dot product. Let $\langle \cdot, \cdot \rangle$ denote an inner product in $\Rb^d$. Then \cref{eq:normed_exp_attention} can be written as
\begin{equation}
    a_{s,j} = \frac{\exp\big( \langle W_q x_s,  W_k x_j\rangle \big)}{\sum_{l=1}^S \exp\big(\langle W_q x_s,  W_k x_l\rangle\big)}
\end{equation}
This normalized exponential is typically referred to in literature as the softmax activation function, defined by
\begin{equation}
    \text{softmax}(z) := \frac{e^{z}}{\sum_{j=1}^N e^{z_j}}.
\end{equation}
With this, the coefficients of the attention are formulated concisely as
\begin{equation}\label{eq:inner_prod_attention}
    a_{s,j} = \text{softmax}\left(\frac{ \langle W_q x_s,  W_k x_j\rangle }{\sqrt{d}}\right),
\end{equation}
where the scaling factor $\frac{1}{\sqrt{d}}$ is added for performance of the network \cite{vaswani2017attention}.

In many applications of transformer networks, a set of self-attention operations are typically performed in parallel, each with a different set of weights $W_q^i$, $W_k^i$, $W_v^i$, for $i = 1, \dots, h$. This is referred to as multi-head attention \cite{vaswani2017attention} and results in parallel projections into a variety of subspaces which are then combined as a linear combination. Multi-head attention is expressed mathematically as 
\begin{equation*}
    \textnormal{MultiHead}(x)=\textnormal{Concat}(y_s^1,\ldots,y_s^h)W_o,
\end{equation*}
where $h$ is the number of heads, and all weights $W_q^i$, $W_k^i$, $W_v^i$, $i=1, \dots, h$, and $W_o$ are trainable.
This multi-head attention mechanism simultaneously grants the model access to different subspaces at different positions in the input sequence. 
\vspace{1em}

\noindent \textbf{Kernel Convolution and Visual Attention Networks}\label{sec:kernel}

\noindent In the seemingly unrelated field of predicting solutions of complex spatiotemporal systems, the use of visual attention has gained popularity, primarily due to the \gls{FNO} \cite{li2020fourier} and the follow-on \gls{AFNO} \cite{guibas2021adaptive}. These methods coalesce in a neural network called FourCastNet \cite{pathak2022fourcastnet} that achieves state-of-the-art prediction accuracy on weather systems. Therein, the task is to generate solutions from a spatiotemporal process that can be written as a partial differential equation. Despite their apparent differences, weather system prediction and natural language processing (as in \cite{vaswani2017attention,li2022quantum}) are unified by underlying sequence data. Indeed, a large variety of tasks involve sequence data, and this is where transformer networks often excel.

The innovation presented in \cite{li2020fourier,guibas2021adaptive,pathak2022fourcastnet} is a perspective that the scaled dot product self-attention mechanism in \cite{vaswani2017attention} can be represented as an integral transform defined by a stationary kernel and treated with the convolution theorem. 
Using \cref{eq:inner_prod_attention}, the attention function may be concisely expressed as
\begin{align}
    y_s = \text{Att}(x_s) := \sum_{s^{\sprime}=1}^{N} a_{s,s^{\sprime}} W_v x_{s^{\sprime}, }
\end{align}
where the coefficients $a_{s,s^{\sprime}}$ are expressed in tensor notation as $a \in \Rb^{N \times N}$. Now self-attention can be described by the action of the asymmetric $d\times d$ matrix-valued kernel $\kappa_{s,s^{\sprime}} := a_{s,s^{\sprime}} W_v$, so that it may be understood as a sum against this kernel. Under the assumption of a stationary kernel $\hat{\kappa}_{s-s^{\sprime}} = \kappa_{s,s^{\sprime}} $, self-attention can be expressed as a global convolution
\begin{equation}\label{eq:sym_kernel_sum}
    \text{Att}(x_s) = \sum_{s^{\sprime}=1}^N  \hat{\kappa}_{s-s^{\sprime}}x_{s^{\sprime}}, \qquad \forall s=1,\ldots,N.
\end{equation}
Finally, leveraging the convolution theorem, one has
\begin{equation} \label{eq:kernel_attention}
    \text{Att}\big(x_s\big) = \calF^{-1}\Big( \calF(\hat{\kappa}) \calF(x) \Big)_s \qquad \forall s = 1, \dots, N,
\end{equation}
where $\calF, \calF^{-1}$ are the forward and inverse Fourier transform, respectively. Note that any transform that satisfies the convolution theorem may be used in place of the Fourier transform. In the \gls{AFNO} network, the kernel is implemented via a complex-valued weight tensor $W:= DFT(\kappa) \in \Cb^{N\times d\times d}$ using the Discrete Fourier Transform (DFT) per token, which acts as a learned weight in the Fourier domain and therefore does not need to be transformed. Unlike the standard perspective of self-attention, \cref{eq:kernel_attention} has a natural extension to quantum circuits where we may similarly learn the kernel in the Fourier domain.

\vspace{1em}

\noindent \textbf{Quantum Machine Learning with Variational Quantum Circuits} 

\noindent \Gls{QML} refers to any of a number of approaches which attempt to exploit the properties of quantum theory to create machine learning models. Here we focus on a method based on \glspl{VQC}. Suppose we are given a data set $\mathcal{S}=\{(x^i,y^i)\}$ consisting of pairs of data points $x^i$ drawn from some data space $\mathcal{R}$ with labels $y^i$ in some set $\mathcal{L}$, and assume there is a function $f:\mathcal{R}\to\mathcal{L}$ which correctly maps a data point to its associated label, i.e. $f(x^i)=y^i$ for all $i$. We wish to approximate $f$ using a \gls{VQC} so that given a new data point $\bar{x}\in\mathcal{R}$, the label $f(\bar{x})\in\mathcal{L}$ can be predicted. Typically $f$ is approximated by a set of variational quantum gates, expressed as parametrized unitary operators $U_\theta$, where $\theta \in [0,2\pi)$ are a set of parameters which are iteratively optimized via any number of optimization algorithms. In what follows, we denote $n$ tensor products of a gate $U$ acting on some Hilbert space $\calH$ by $U^{\otimes n}$.

The construction of the \gls{VQC} begins with an embedding of the data into a Hilbert space, and this alone is a nontrivial step, as the choice of embedding can have profound consequences on the trainability of the model \cite{schuld2021machine,schuld2021supervised} (see also \cite{laRose2020} for robustness results). Possibly the most obvious embedding is the basis embedding, where the binary representation $B(x)$ of the data point $x$ is embedded into a Hilbert space by mapping $x$ to the computational basis state $\ket{B(x)}$. However, this approach is extremely resource inefficient, as the necessary number of qubits is equivalent to the number of bits in the binary representation. On the opposite end of the spectrum is the amplitude embedding, where the data features are encoded in the coefficients of a quantum state. Since there are $2^n$ coefficients in an $n$-qubit quantum state, the required number of qubits to implement this embedding is $O(\log(n))$, where $n$ is the number of features in $x$. However, the problem of preparing an arbitrary state is thought to be hard, so that the actual implementation of this embedding is difficult to perform in practice. Somewhere in the middle is the so-called angle embedding, where the data features are encoded in the angles of $n$ rotation operators which are then applied to the $\ket{0}^{\otimes n}$ state. The angle embedding therefore requires $O(n)$ qubits. Another option is to train the embedding to maximally separate data classes in a Hilbert space as suggested by Lloyd et al. \cite{lloyd2020quantum}. Finally we also note a more recent resource efficient embedding for image data called the QPIXL embedding \cite{amankwah2022quantum}, which simultaneously acts as a pixel value embedding and a positional encoder by storing the location and value of a pixel in a tensor product. For now, we will assume that an embedding $\mathcal{E}:\mathcal{R}\to\mathcal{U}(\mathcal{H})$ has been chosen, where $\calU(\calH)$ is the space of unitary operators on $\mathcal{H}$.

Typically, the next step is to apply a parameterized unitary operator $U_\theta$, which can have many forms. Generally, this operation should entangle the data qubits together to allow information to be shared, thereby enhancing expressivity. Note that the QFT, although unparameterized, was shown to have this property \cite{mastriani2021quantum}. The final step in the variational circuit is to make a measurement with respect to an observable $\calO$, most generally belonging to the set of \glspl{POVM}, which is then used to predict the label associated to the given data point. The expected value of this circuit is given by
\begin{equation} \label{eq:estimate}
h_{\theta}(x)=\tr\left[U_\theta(\mathcal{E}(x)\ket{0}\!\!\bra{0}\mathcal{E}(x)^\dagger) U^\dagger_\theta\mathcal{O}\right],
\end{equation}
We interpret \cref{eq:estimate} as an estimate of the function $f$, and our goal is to minimize the error in this estimation.

The \gls{VQC} is trained using any of a number of optimization schemes. Early optimization methods include the Nelder-Mead \cite{peruzzo2014variational} and other zeroth-order or direct search methods; however, these have largely been replaced by methods that at least approximate the gradient of the variational circuit with respect to its trainable parameters. Notably, the parameter-shift rule and its stochastic variant are first-order optimization methods which allow us to recover the exact analytic gradient by simply running the circuit with the parameters shifted up and down \cite{mitarai2018,schuld2019,banchi2021measuring}; however, it requires $O(3mr)$ evaluations of the circuit, where $m$ is the number of variational parameters and $r$ is the number of shots used to approximate the expectation. A popular alternative is the \gls{SPSA} method \cite{spall1998overview}, which is a quasi-first order method that approximates the gradient via a random simultaneous shift of \textit{all} parameters, and thus only requires $O(r)$ evaluations of the circuit. Variants of these methods incorporate the quantum natural gradient via the quantum Fisher information matrix, as in QN-SPSA \cite{gacon2021simultaneous}, incorporate approximate second-order (Hessian) information as in 2SPSA \cite{spall2000adaptive} or L-BFGS \cite{berahas2016multi}, or incorporate well-established classical machine learning techniques such as adaptive learning rates or momentum \cite{kingma2014adam}.

\vspace{1em}
\noindent \textbf{SASQuaTCh: A Quantum Fourier Vision Transformer Circuit} 

\begin{figure*}[t]
    \centering
\[
\input{QFVT}
\]
    \caption{The \gls{SASQuaTCh} circuit applied to the context of classification, where a single readout qubit is used to classify a two-class problem.}
    \label{fig:QFVT}
\end{figure*}
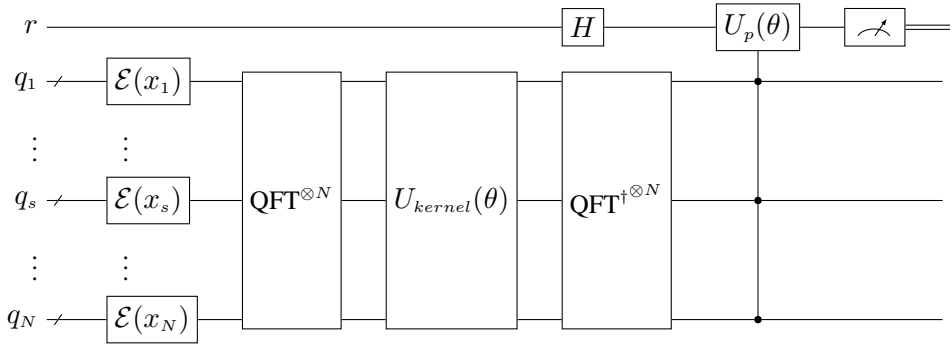

\noindent We now present the construction of our quantum vision transformer. The variational quantum circuit is designed to 1) incorporate sequence data, 2) perform the self-attention mechanism \textit{inside} the quantum circuit, and 3) provide a problem specific readout. Utilizing the perspective provided by the \gls{AFNO} network, self-attention is applied using a convolution against a kernel. Our image sequence is represented as a compact tensor $x = [x_1, x_2, \dots, x_s, \dots, x_N]^\T \in \Rb^{N \times d}$. Let $\ket{\psi_s}, s=1, \dots, N$ denote an encoded quantum state of classical sequence element $x_s$, and let $\ket{\psi} = \ket{\psi_1} \otimes \ket{\psi_2} \otimes \cdots \otimes \ket{\psi_N}$ denote the entire sequence. The encoded quantum state of each sequence element is of the form
\begin{equation}
    \ket{\psi_s} := \calE(x_s) \ket{0}^{\otimes \hat{d}}
\end{equation}
where $\hat{d} \leq d$ is the number of qubits used to encode the $d$-dimensional input vector $x_s$.
As described earlier, the problem of embedding classical data into a quantum circuit is nontrivial, although there are a number of well-established approaches, including the angle and amplitude embeddings. Inspired by the position embedding used by QPIXL, our approach stores positional information by means of the amplitude embedding. Specifically, a $2^m\times 2^m$ grayscale image is represented by the quantum state
\begin{equation} \label{eq:amplitude_embed}
    \ket{\psi_s} = \frac{1}{\sqrt{2^m}}\sum_{k=0}^{2^{2m}-1} x_{s,k} \ket{k},
\end{equation}
where $x_{s,k}$ is the grayscale value of the $k$-th pixel of the sequence element $s$, and $\ket{k}$ is the $k$-th binary basis element of the Hilbert space. The extension to non-square images is accomplished by padding. There may exist alternative approaches which efficiently embed the entire sequence into qubits, but we leave this to future work.

\begin{figure*}[t]
    \centering
\[
\input{perceptron}
\]
    \caption{The variational ansatz $U_{p}(\theta)$ used in the \gls{SASQuaTCh} circuit to perform classification tasks. The readout qubit $\ket{r}$ is conditionally rotated controlling on the data qubits, and the parameters $\theta_i$, $i=1,\dots,4N$ are updated as a part of the optimization routine.}
    \label{fig:perceptron}
\end{figure*}
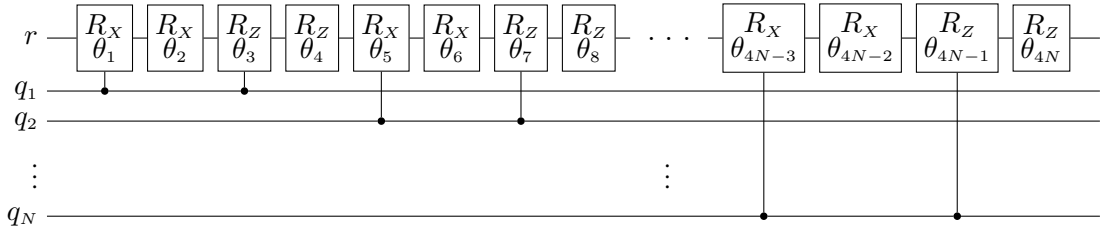

Our approach is named the \glsfirst{SASQuaTCh} and is depicted in \cref{fig:QFVT}. A \gls{QFT} is applied to each encoded sequence element, followed by a variational ansatz $U_{kernel}(\theta)$ applied to \textit{all} qubits to perform channel mixing. Details of this variational circuit are depicted in \cref{fig:entangling_unitary}. This is followed by an inverse \gls{QFT} on each sequence element. \gls{SASQuaTCh} is applicable to a variety of machine learning tasks, however here we restrict our attention to classification problems and thus we apply a parameterized unitary $U_p(\theta)$ to the readout qubit controlled off of the data qubits. This has the effect of transferring information from the data register to the readout register, where a measurement is then made to obtain a prediction. The ansatz of $U_p(\theta)$ is depicted in \cref{fig:perceptron} and is easily generalized to more than one readout qubit e.g. for a many-class classification problem where additional readout qubits can be used to form a binary expansion of the number of classes.

\gls{SASQuaTCh} makes use of a kernel-based attention mechanism
\begin{equation}
    \ket{\phi} := QFT^{\dagger^{\otimes N}} U_{kernel}(\theta) QFT^{\otimes N} \ket{\psi}
\end{equation}
which is directly analogous to \cref{eq:kernel_attention} with matrix-valued kernel $\calF(\kappa) =  U_{kernel}(\theta)$ in the Fourier domain. This provides a straightforward implementation of the self-attention mechanism in a quantum circuit as well as design insight on the choice of variational ansatz $U_{kernel}$. In the context of the kernel integral perspective of self-attention, the variational ansatz $U_{kernel}(\theta)$ serves as a trainable kernel which acts in Fourier space to perform channel mixing. The 
variational ansatz in \gls{SASQuaTCh} should be chosen to mutually entangle the Fourier space representation of the embedded qubit sequence. For this reason, we make use of Pennylane's built-in \texttt{StronglyEntanglingLayers} operation \cite{bergholm2018pennylane} which is based on the circuit-centric classifier design in \cite{schuld2020circuit}. There are likely many suitable choices for $U_{kernel}(\theta)$, and we leave this exploration to future work.

The readout qubits also serve a critical role in the trainability of the overall circuit, since the objective function
\begin{equation} \label{eq:loss}
    L(\theta) := \calD\Big(y_{label}, \langle \mathcal{O} \rangle \Big)
\end{equation}
is defined using the expected value of the readout register with respect to an observable $\mathcal{O}$. Here, $y_{label}$ is the set of true labels, $\calD$ is a metric such as the $l_2$ norm or the cross-entropy, and the expectation is taken on the readout register. The information stored in this readout qubit is the result of the variational ansatz $U_p(\theta)$, which while independently constructed, has similarities to the quantum perceptron introduced in \cite{maronese2022quantum}. This circuit is used to rotate the readout qubit conditioned on data qubits, and the uncontrolled rotation gates serve as a bias term analogous to classical machine learning. The resulting \gls{SASQuaTCh} architecture accomplishes the desired goals of the design of a quantum transformer circuit and is applied in the sequel to a classification task wherein the input data is a sequence. However, as previously noted, we expect this architecture to be applicable to \textit{any} problem involving sequence data.

\begin{figure*}[t]
    \centering
\[
\input{entangling_unitary}
\]
    \caption{The variational ansatz $U_{kernel}(\theta)$ used in the \gls{SASQuaTCh} circuit. For $n$ qubits and $l$ layers, the parameters $\theta = [\alpha_1^1, \beta_1^1, \gamma_1^1, \dots, \alpha_n^l, \beta_n^l, \gamma_n^l]$ are updated as a part of the optimization routine.}
    \label{fig:entangling_unitary}
\end{figure*}
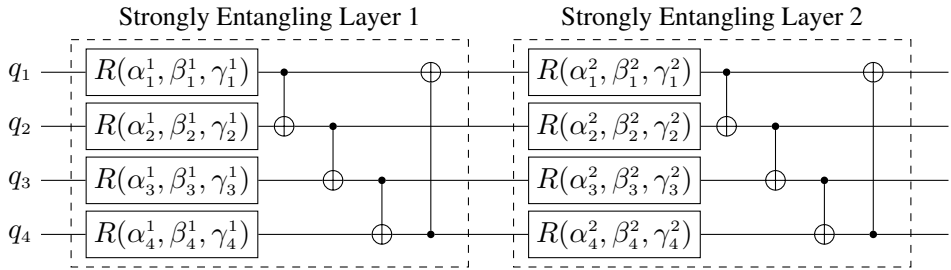

\vspace{1em}
\noindent \textbf{Sequential Quantum Vision Transformers}

\noindent The prevalent approach in machine learning literature is one of deep learning, where layers of a given operation are performed sequentially. These layers are typically separated by nonlinearities referred to as nonlinear activation functions, which both prevent sequential linear operations from reducing to a single linear operation and critically provide flexibility so that the neural network can more easily approximate the desired functional mapping. 

As suggested by the name of our approach, the \gls{SASQuaTCh} architecture enables deep layering of the self-attention mechanism within a single quantum circuit. In contrast to approaches such as \cite{li2022quantum}, one does not need to classically read out the results of one QSAL circuit before preparing the next one. Instead, as in \cite{guibas2021adaptive}, we may refer to \cref{fig:QFVT} as a single layer, and sequentially stack layers to yield a deep \gls{SASQuaTCh} circuit.

Note that the authors in \cite{guibas2021adaptive} use nonlinearities as part of their kernel to form a \gls{MLP}. However in contast to the classical ML literature, the need for nonlinearities in \gls{QML} is still largely an open research topic. Several nonlinearities for quantum circuits have been proposed in \cite{gili2023introducing,holmes2023nonlinear,maronese2022quantum}, such as versions of the rectified linear unit (ReLU), the \gls{RUS} nonlinearity, and a quantum perceptron. We leave experimentation with nonlinearities for future research.

\vspace{1em}
\noindent \textbf{Image Classification Experiments} 

\noindent Inspired by the development of classical vision transformers, here we validate the efficacy of the \gls{SASQuaTCh} architecture on image classification tasks. We note however that this framework is generic and can be similarly applied to the vast array of machine learning tasks involving sequence data, such as natural language processing and time-series prediction.

Our first experiment involves a simple synthetic dataset and investigates if \gls{SASQuaTCh} can find a meaningful feature space for classification. The task in this case is to classify the line in the noisy image as horizontal or vertical. From this experiment we move to the harder task of classifying handwritten digits from the MNIST dataset \cite{lecun2010mnist}. We further split this second experiment into an easier subcase of classifying the handwritten digits 1 and 3, and a significantly harder subcase of classifying the handwritten digits 3 and 8. Our experiments test a variety of hybrid network architectures, including \gls{SASQuaTCh} with a variable number $l$ of \texttt{StronglyEntanglingLayers}.

In all of our experiments, we first perform a classical patch and position embedding layer as in \cite{guibas2021adaptive}, which yields a sequence of patches that can be treated with quantum self-attention. These patches are represented by vectors of embedding dimension $\varepsilon$, and are encoded by either an angle encoding or an amplitude encoding. A single readout qubit is then passed to a single linear output layer and ultimately provides an expectation of the loss function \cref{eq:loss}, which we take as the mean absolute error or $l_1$ loss. This hybrid scheme is built using PennyLane \cite{bergholm2018pennylane}, where we use PennyLane's \texttt{TorchLayer} function to yield a PyTorch \cite{Paszke2019PyTorch} module which can integrate with the classical portions of our hybrid model. Finally, since the architecture is trained completely in simulation, optimization is performed using the ADAM optimizer \cite{kingma2014adam}.

\subsection*{Synthetic Line Classification}\label{ssec:synthetic_lines}

\begin{figure}[t]
\centering
\includegraphics[width=.24\textwidth,trim={2.5cm 1.25cm 2.5cm 1.25cm},clip]{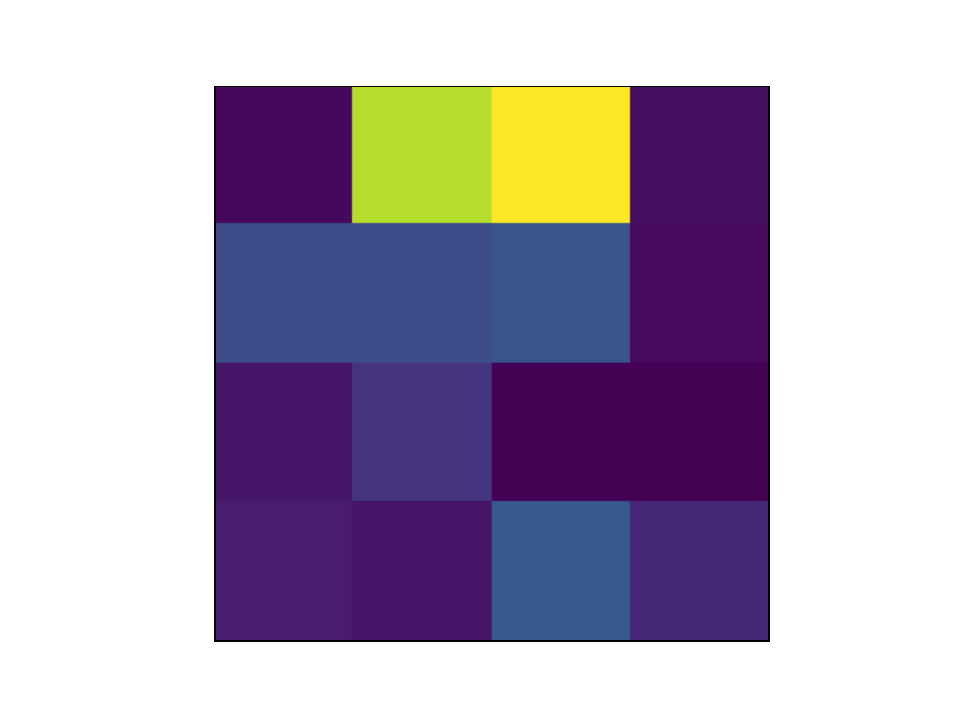}
\includegraphics[width=.24\textwidth,trim={2.5cm 1.25cm 2.5cm 1.25cm},clip]{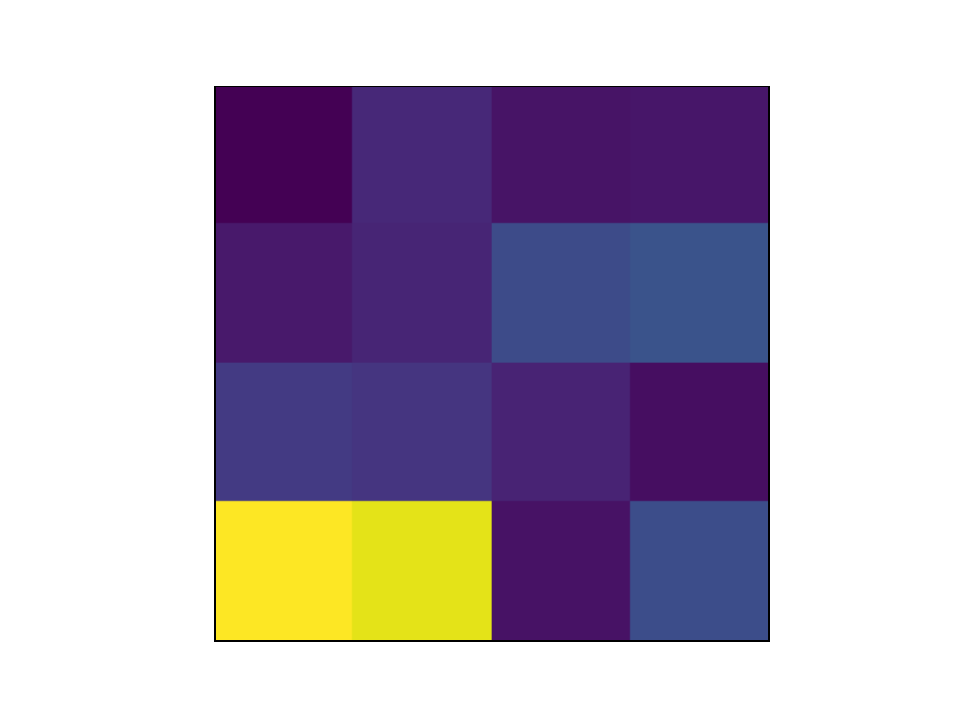}
\includegraphics[width=.24\textwidth,trim={2.5cm 1.25cm 2.5cm 1.25cm},clip]{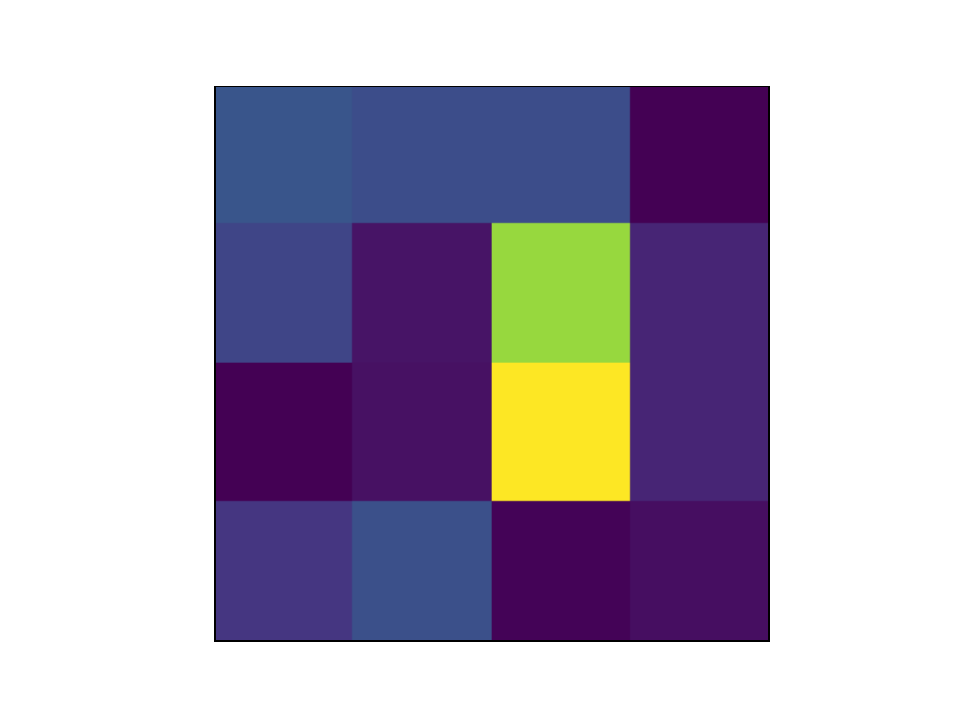}
\includegraphics[width=.24\textwidth,trim={2.5cm 1.25cm 2.5cm 1.25cm},clip]{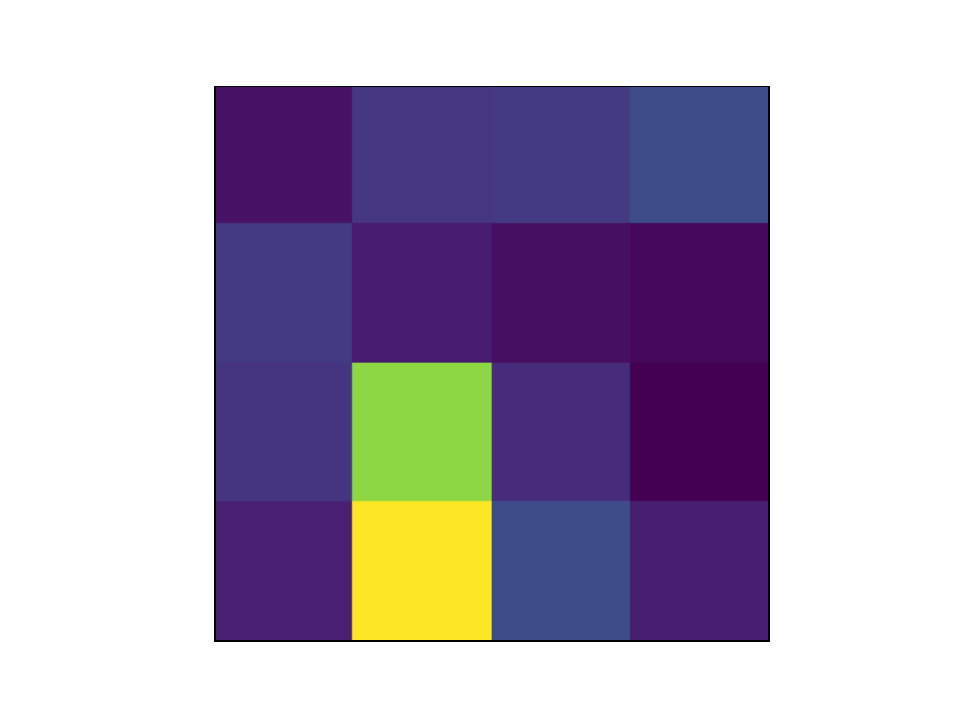}
\caption{Examples of synthetically generated noisy images with horizontal or vertical lines. The images are greyscale, with color added for visualization purposes only.}
\label{fig:SynData}
\end{figure}

For the synthetic line classification experiment we generate noisy synthetic $4\times4$ grayscale images where two adjacent pixels define a line, which can either be horizontal or vertical, as in this Qiskit demo\footnote{\url{https://qiskit-community.github.io/qiskit-machine-learning/tutorials/11_quantum_convolutional_neural_networks.html}}. In our implementation we set the line to have a magnitude of $0.75$ then add uniformly sampled random noise in the range of $[0, 0.25]$ to the entire image. This gives us a normalized image without any additional processing. Examples of the generated images are shown in \cref{fig:SynData}.

For our patch and position embedding, we choose a patch size of $2\times2$ with an embedding dimension $\varepsilon = 4$. Note that in this experiment we leave the embedding untrained, which in essence acts as a random projection from image space to patch sequence space. This leads to a sequence of four patches with each patch represented by a linear projection to a four dimensional vector. We note that in this case our classical embedding has no loss of information of the original image. The vectors representing each patch are then encoded in the quantum circuit using angle embedding, which results in a total of 17 qubits and 144 trainable parameters.

\begin{figure}[t]
    \centering
    \includegraphics[width=.49\textwidth]{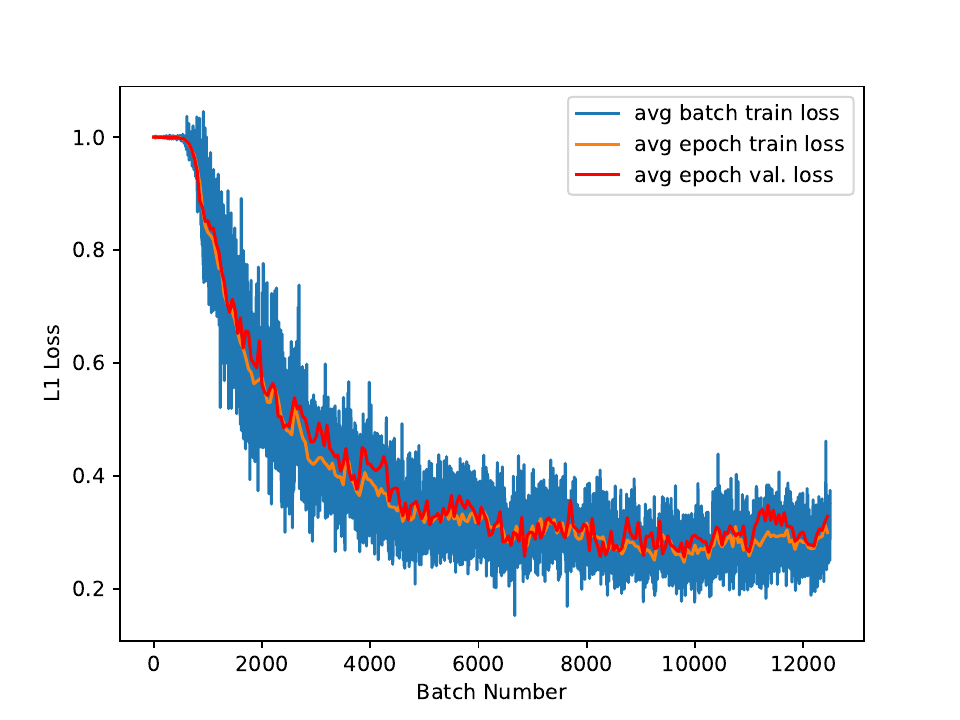}
    \includegraphics[width=.49\textwidth]{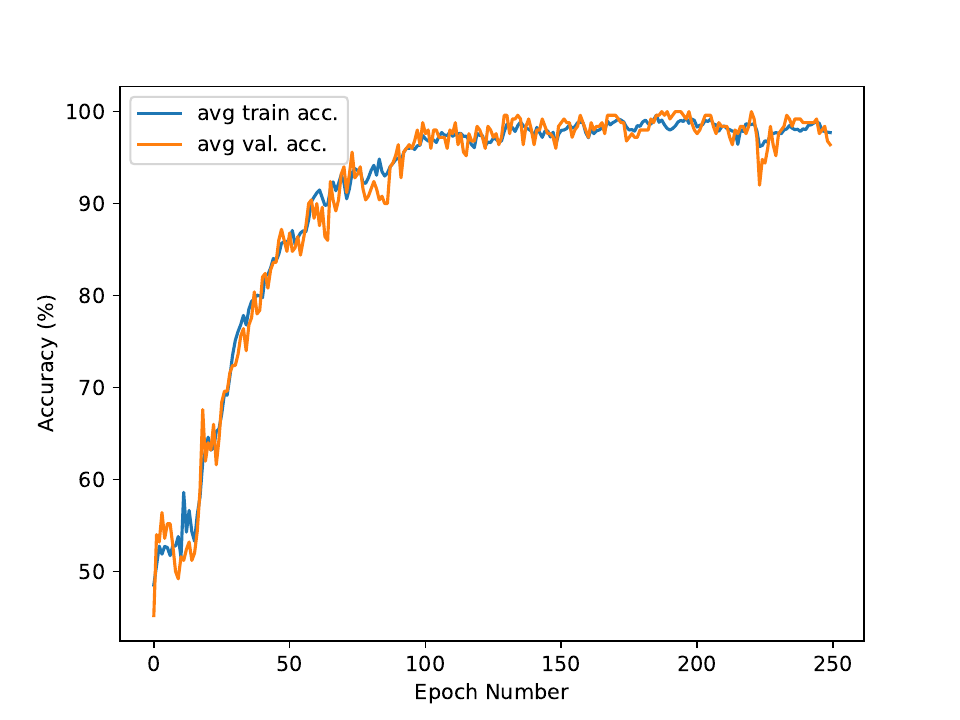}
    \caption{Loss function as a function of training time (epochs), both training and validation losses are depicted for the synthetic data on simulated hardware. 
    }
    \label{fig:SynLossvsEpoch}
\end{figure}

The training dataset is composed of a total of 500 images, 250 of each class, and our hybrid network is trained for 100 epochs with an initial learning rate of $0.001$. The learning curves for an example training run are shown in \cref{fig:SynLossvsEpoch}. We compare the \gls{SASQuaTCh} circuit as detailed above to an identical circuit with only the \gls{QFT} and inverse \gls{QFT} removed, and to a baseline where the quantum circuit only includes an angle encoding and a perceptron. We report a summary of the performance of each method in \cref{tab:ResSimSyn}, in which the average training and validation accuracies from 5 runs are reported, where each run has a random parameter initialization. The comparison clearly demonstrates the value of including the \gls{QFT} operations necessary to perform our quantum kernel-based self-attention mechanism. In the case of the baseline and the \gls{SASQuaTCh} circuit without \gls{QFT} operations, the results indicate that these models perform marginally better than a random guess, while \gls{SASQuaTCh} successfully classifies the lines in the images with high accuracy.

\begin{table}[h]
    \caption{Training and validation accuracy results on the synthetically generated line classification image data. Accuracy values are given as a percentage, and the values in parentheses represent one standard deviation.}
    \centering
    \begin{tabular}{c || c c c}
                            & Train Acc. & Validation Acc. \\ \hline\hline
        SASQuaTCh           & 96.84 (2.87) & 96.8 (4.49) \\ 
        SASQuaTCh (no QFT)  & 56.04 (4.40) & 53.2 (2.71) \\
        Quantum Base.       & 58.48 (2.39) & 51.6 (5.43)
    \end{tabular}
    \label{tab:ResSimSyn}
\end{table}

\subsection*{Classifying Handwritten Digits}\label{ssec:SimMNIST}

\begin{figure}[t]
    \centering
    \includegraphics[width=.49\textwidth]{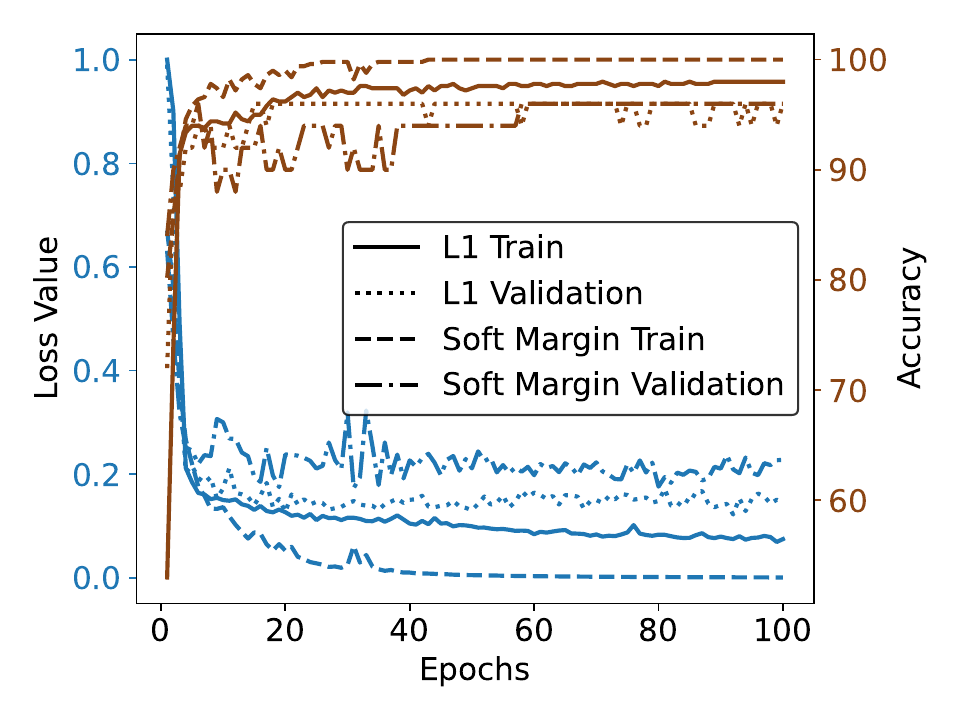}
    \caption{Loss (blue) and accuracy(orange) curves of the $\{1,3\}$ handwritten digit pair classification experiment over 100 epochs. The line styles solid, dashed, etc. indicate which loss function and sub dataset was used.}
    \label{fig:1vs3loss_comparison}
\end{figure}

Our second experiment uses the very common MNIST machine learning dataset composed of handwritten digits ranging from 0 to 9. For simplicity, we consider only binary classification using a single readout qubit, however we note that the \gls{SASQuaTCh} architecture can be extended to multi-class tasks by performing multiple quantum perceptrons between the data qubits and a set of readout qubits. The difficulty of classifying an arbitrary pair of handwritten digits can vary substantially, and here we select a slightly easier classification pair $\{1,3\}$, and perhaps the most challenging classification pair $\{3,8\}$. 

Both of these experiments present a significantly more challenging task than the synthetic line classification task. Whereas the images in the synthetic line classification task were $4\times4$ pixels, the images in the MNIST dataset are $28 \times 28$ pixels with much more complex features. Using the angle encoding in the case of MNIST images would require many more qubits, however the flexibility of the \gls{SASQuaTCh} architecture enables the use of the far more qubit efficient amplitude encoding. Additionally, since the amplitude encoding of an image requires only logarithmic qubits, it is prudent to use a classical patch and position embedding with a larger patch size, which we take nominally as $16\times16$ with an embedding dimension $\varepsilon=4$. With this combination of patch size, embedding dimension, and qubit encoding, a patch sequence generated from an MNIST image can be encoded using only eight qubits. Thus our entire \gls{SASQuaTCh} circuit with one readout qubit requires only nine qubits to perform handwritten digit pair classification on the MNIST image dataset.

For the  $\{1,3\}$ digit pair experiment the MNIST image dataset is uniformly sampled into 1000 images for the training set and 100 images for the validation set. All 2145 images from the test set are used. Our hybrid network is trained in simulation for 200 epochs, with an example of the training and validation curves depicted in \cref{fig:1vs3loss_comparison}. Here we also experiment with the choice of loss function between the soft margin loss and the $l_1$ loss. The trained hybrid network was then evaluated using the \texttt{ibm\_fez} QPU. The experiments yielded an accuracy of 96.8\% on a blind test set, averaged over 8192 shots.

\begin{figure}[t]
    \centering
    \includegraphics[width=0.49\textwidth]{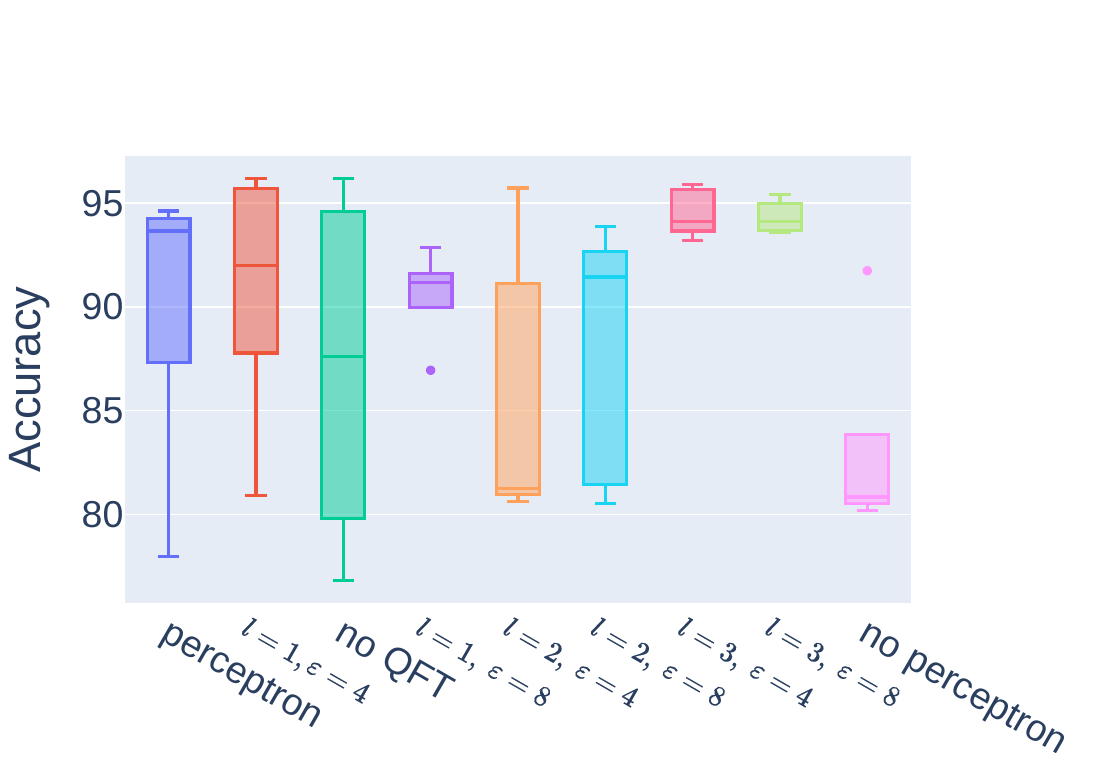}
    \caption{Test accuracy ablation study on the $\{3,8\}$ handwritten digit classification task using 9 cases involving the \gls{SASQuaTCh} architecture, where $\varepsilon$ denotes embedding dimension and $l$ denotes number of \texttt{StronglyEntanglingLayers}: 1) the perceptron alone with $\varepsilon=4$; 2) \gls{SASQuaTCh} with $\varepsilon=4$, $l=1$; 3) \gls{SASQuaTCh} without the \gls{QFT} operations, $\varepsilon=4$, and $l=1$; 4) \gls{SASQuaTCh} with $\varepsilon=8$ and $l=1$; 5) \gls{SASQuaTCh} with $\varepsilon=4$ and $l=2$; 6) \gls{SASQuaTCh} with $\varepsilon=8$ and $l=2$; 7) \gls{SASQuaTCh} with $\varepsilon=8$ and $l=3$; 8) \gls{SASQuaTCh} with $\varepsilon=8$, $l=3$; and 9) \gls{SASQuaTCh} without a perceptron, with $\varepsilon=4$ and $l=1$.}
    \label{fig:3vs8ablation}
\end{figure}

For the substantially harder $\{3,8\}$ digit pair experiment the MNIST image dataset is similarly uniformly sampled into 1000 training images and 100 validation images. Again, all 1984 test images are used. In this experiment we ran an ablation study on a variety of cases as depicted in \cref{fig:3vs8ablation}, including variations of the embedded dimension $\varepsilon$ of each patch. Each case was trained for 200 epochs from 5 random parameter initializations, and the figure depicts accuracy over the test set. The figure depicts large variances for several cases, including \gls{SASQuaTCh} with no \gls{QFT} operations, and very small variances for other cases such as using an embedding dimension $\varepsilon=8$ and three \texttt{StronglyEntanglingLayers}. These results demonstrate the ability of \gls{SASQuaTCh} to have strong performance in image classification tasks. 

We note a strong correlation between the variance of several cases and both hyperparameters $\varepsilon$ and $l$, suggesting a sensitivity to initial conditions of the trainability of the \gls{VQC} which depends on hyperparameters. The cases that have high variance indicate the presence of local minima or potentially barren plateaus in the optimization landscape, and furthermore that the presence of these local minima may change with $l$. These cases can be easily distinguished by calculating the variance of the loss function. In any case, this apparent issue appears to be mitigated by layering and proper choice of hyperparameters in the case of the \gls{SASQuaTCh} architecture.

The results indicate substantial value in layering kernel operations inside the \gls{QFT}, which is analogous to deep layers of self-attention in the classical context. In contrast to \gls{QSANN}, the deep layering of the \gls{SASQuaTCh} architecture does not require repeated measurement operations and re-encoding of classical self-attention operations. Instead adding depth using the \gls{SASQuaTCh} architecture is as simple as extending the quantum circuit with additional \texttt{StronglyEntanglingLayers} inside the \gls{QFT}.

\begin{figure}[t]
    \centering
    \includegraphics[width=.5\textwidth]{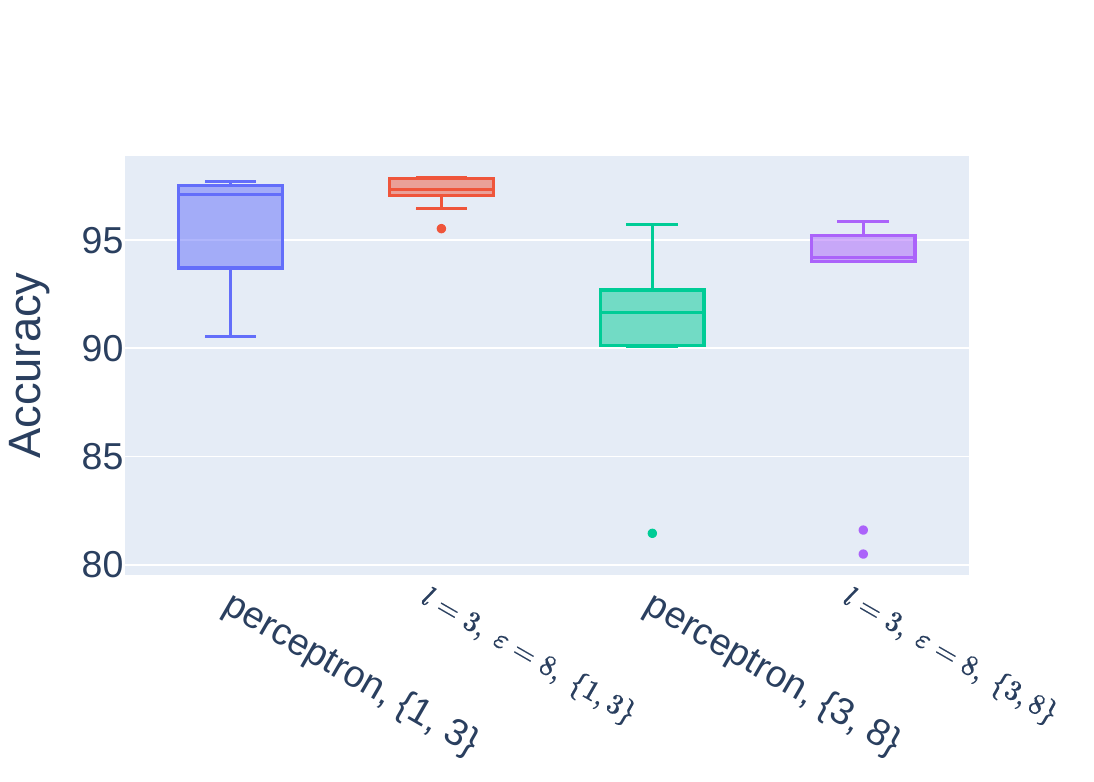}
    \caption{Test accuracy over 10 independent runs involving the \gls{SASQuaTCh} architecture comparing the best \gls{SASQuaTCh} hyperparameters against the optimized perceptron, where $\varepsilon$ denotes embedding dimension and $l$ denotes number of \texttt{StronglyEntanglingLayers}: 1) the perceptron for the $\{1,3\}$ task with $\varepsilon=4$, 2) \gls{SASQuaTCh} with $\varepsilon=8$ and $l=3$ for the $\{1,3\}$ task, 3) the perceptron for the $\{3,8\}$ task with $\varepsilon=4$, 4) \gls{SASQuaTCh} with $\varepsilon=8$ and $l=3$ for the $\{3,8\}$ task.}
    \label{fig:1vs3_3vs8}
\end{figure}    

\section*{Discussion \& Conclusion}\label{sec:conclusion}

This work serves as a novel approach to applying the self-attention mechanism as popularized by the widespread success of transformer neural networks entirely inside of a quantum circuit. A recent perspective of the self-attention mechanism as a kernel integral is leveraged in the design of a \gls{VQC} that effectively applies the self-attention mechanism without the need to apply lossy measurements between layers. We investigate the design of an appropriate kernel ansatz and in the context of classification tasks, apply a variational ansatz to transfer relevant information to a readout qubit. The resulting method benefits from logarithmic qubit complexity, logarithmic parameter complexity, and logarithmic gate complexity with respect to the size of the data. To the best knowledge of the authors, this work is the first to produce a quantum vision transformer in this way.

We note in the context of \cite{bermejo2024quantum} that several subsets of the MNIST handwritten digit classification tasks have been shown to be solvable with encoding alone. However, the selected digit pairs in our experiments do not belong to this set, and furthermore the digit pair $\{3,8\}$ is likely the most challenging two-digit sub-task. While this is a relatively standard dataset for benchmarking performance, the synthetic line classification task appears to be more challenging due to the likelihood of the line being segmented into separate patches. Thus in this experiment, the mixing of channels through the self-attention mechanism is likely to be a larger contributor to the overall success of the approach. Future work will expand on these results by tackling more challenging classification tasks such as CIFAR-10 and other multi-class tasks. At present, our software implementation is limited by a 32-qubit python constraint \cite{Our_comment}.

We have introduced the \gls{SASQuaTCh} architecture, a novel quantum transformer model which implements self-attention in a fully quantum setting by utilizing the \gls{QFT}. There are many directions to explore that will likely lead to improvements in the model. One such direction is a study of the relationship between the success of the model and the data embedding used in the patching procedure; although, this is likely problem dependent, so that we do not expect one particular embedding to triumph in every context. Similarly, the structure of the channel-mixer $U_{kernel}$ and the perceptron $U_p$ can be experimented with and may have a most successful structure that is problem-dependent. Yet another option is to explore the use of \gls{SASQuaTCh} on non-image sequence data such as dynamical systems, natural language, or time series data. Finally, one can explore the use of nonlinearities inside the quantum circuit sandwiched between each consecutive layer as a means to increase expressibility.

\section*{Methods}

\noindent \textbf{Data Embedding Schemes and Complexity Analysis}

For a large variety of VQCs, the scaling of the number of qubits, gate depth, and number of parameters has a strong dependence on the data embedding scheme. We find this to be true also in the case of the SASQuaTCh architecture. Consider as before a grayscale image of shape $2^m \times 2^m$. Furthermore, let $q$ denote the number of qubits, $g$ denote the number of gates, and $p$ denote the number of variational parameters. 

\textbf{Angle Embedding: } With the angle embedding
\begin{equation}
    \ket{\psi_s} = \bigotimes_{k=0}^{4^{m}-1} U(x_{s,k}) \ket{0}
\end{equation}
one requires $q=2^{2m}=4^m$ qubits, where $U(x_{s,k}) \in \{R_x, R_y, R_z\}$ is some rotation gate. Thus one has linear qubit complexity in the number of pixels $q = \calO(4^{m})$. However, assuming the hardware implementation allows parallel gate operations, one has a constant gate depth embedding $g=\calO(1)$.

The use of visual attention is based on developing a sequence of patches from the original image. Assume the $2^m \times 2^m$ image is split into a sequence of $N$ patches of shape $2^v \times 2^v$. If each patch is embedded by angle embedding then one has qubit complexity $q = \calO(N 4^{v})$ with again constant gate complexity $g = \calO(1)$. Applying the \gls{SASQuaTCh} architecture preserves the number of qubits up through the perceptron, which adds $\calO(\log c)$ qubits for a $c$-class classification task. Thus the total number of qubits becomes $q = \calO(N 4^{v} + \log c)$.

The parameters of this architecture live inside the variational kernel composed of $l$ \texttt{StronglyEntanglingLayers} as well as the perceptron. Each \texttt{StronglyEntanglingLayer} has $3q$ parameters, thus the number of parameters in the kernel is $p_{\text{kernel}} = 3ql = \calO(3 lN 4^{v})$. The perceptron uses four parameters for each data qubit, however it is not completely understood how the perceptron parameter scaling depends on the number of readout qubits $r = \log c$ for a $c$-class classification problem. We consider as the worst case the naive implementation of one perceptron per readout qubit. Thus $p_p = \calO(4q\log c) = \calO(4 N 4^{v} \log c)$. Combining these two contributions, the total parameter complexity under angle embedding is $p = \calO\big(N 4^{v} (l + \log c) \big)$.

Assuming the efficient approximation \cite{hales2000improved} the \gls{QFT} and its inverse each contribute a number of gates scaling with data qubits $g_{QFT} = \calO(q \log q) = \calO\big(N 4^v \log (N 4^v)\big)$. With parallel gates the $l$ \texttt{StronglyEntanglingLayers} contribute $g_{\text{kernel}} = \calO(q) = \calO(N4^v)$, and each perceptron contributes $g_{p} = \calO(4q) = \calO(4 N 4^v)$. Thus assuming the worst case naive implementation of one perceptron per readout qubit, one has $g_p = \calO(4N 4^v \log c)$. Ultimately, the overall circuit has a gate complexity of $g = \calO\big(N 4^v\log (cN 4^v)\big)$.

\textbf{Patch Amplitude Embedding:} With the amplitude embedding in \cref{eq:amplitude_embed} applied to the entire grayscale image, one requires logarithmic qubits $q = \calO(2m)$, however after the image is broken into a sequence of $N$ patches, one has $q = \calO(2Nv)$. As with the angle embedding, the parameter scaling is largely inherited from the embedding scheme, thus in the present case one has linear growth in the number of qubits or logarithmic parameter growth with respect to the image size $p = \calO(3ql + 4q\log c) = \calO\big(Nv(l + \log c) \big)$.

While the number of gates of the amplitude embedding is traditionally understood to scale exponentially with the number of qubits \cite{plesch2011quantum}, a recent approximation method achieve polynomial scaling with the number of qubits \cite{nakaji2022approximate}. In the worst case, the exponential gate complexity of the amplitude embedding yields total gate complexity of $g =g_{\text{AE}}+g_{\text{QFT}} + g_{\text{kernel}} + g_p = \calO\big(q + q\log(cq)\big) = \calO\big(4^{Nv} + Nv\log(cNv)\big)$, which scales linearly with patch size. However, using the approximation scheme \cite{nakaji2022approximate} yields $g = \calO\Big(Nv\big(b+\log(cNv)\big)\Big)$ or logarithmic gate growth with respect to patch size, where $b$ is the depth of the approximation.

\vspace{1em}
\noindent \textbf{Continuous Kernel Integral}

\noindent As an aside, the kernel summation can be extended to the continuous integral setting \cite{li2020fourier}. While this may seem unnecessary in the context of finite Hilbert spaces as in qubit systems, it is particularly relevant to uncountably infinite Hilbert spaces as in continuous variable quantum computing. Let $\calD \subset \Rb^2$ be a bounded domain and let $\calH_d = \{ f: \calD \rightarrow \Rb^d \}$ be a Hilbert space of functions from $\calD$ to $\Rb^d$, so that $X\in \calH_d$ can be viewed as a spatial function. Define the integral transform $\calK: \calH_d \rightarrow \calH_d$ by
\begin{equation}\label{eq:continuous_int_transform}
    \calK\big(X\big)(s) = \int_D \kappa(s, s^{\sprime}) X(s^{\sprime}) \rd s^{\sprime} \qquad \forall s \in D,
\end{equation}
where $\kappa: D \times D \rightarrow \Rb^{d\times d}$ is a continuous kernel function. If $\kappa$ is once again assumed to be stationary $\kappa(s,s^{\sprime}) = \kappa(s-s^{\sprime})$, then \cref{eq:continuous_int_transform} becomes a global convolution
\begin{equation}
    \calK\big(X\big)(s) = \int_D \kappa(s - s^{\sprime}) X(s^{\sprime}) \rd s^{\sprime} = \big( \kappa \ast X \big) (s), \; \forall s \in D.
\end{equation}
Leveraging the convolution theorem in a similar fashion, one has
\begin{equation} \label{eq:continuous_kernel_attention}
    \calK\big(X\big)(s) = \calF^{-1}\Big( \calF(\kappa) \calF(X) \Big)(s) \qquad \forall s \in D,
\end{equation}
and the continuous integral transform \cref{eq:continuous_kernel_attention} then performs self-attention on continuous spaces. This continuous perspective is leveraged in works such as \cite{pathak2022fourcastnet,salvi2022neural} to obtain predictions of PDE systems which are invariant under the resolution of the spatial discretization, a property that has tremendous utility in the context of predicting solutions to spatiotemporal physical systems.

\vspace{1em}


\newcommand{\BIBdecl}{\setlength{\itemsep}{0em}}
\bibliographystyle{ieeetran}

\balance
\bibliography{References}

\section*{Acknowledgments}

MLL and ENE were supported by the Department of Defense through the SMART Scholarship program and the SMART SEED grant at \href{www.smartscholarship.org}{www.smartscholarship.org}. This work was sponsored by the Naval Innovative Science \& Engineering (NISE) program at NSWC PCD.

Simulations were performed using Pennylane's python packages \cite{bergholm2018pennylane}, and example results were achieved using IBM Q machines. We acknowledge use of the IBM~Q for this work. The views expressed are those of the authors and do not reflect the official policy or position of IBM or the IBM~Q team. 

This document has been approved for public release; distribution is unlimited.

\section*{Author Contributions}

\noindent ENE first conceived the idea, wrote the initial code, helped design experiments, and was the primary paper writer. MC was the lead code developer, helped design the experiments, and wrote the first draft of the experiments section. ZPB wrote large sections of the paper, contributed the use of the \texttt{StronglyEntanglingLayer}, and contributed to the mathematical formalism. MLL helped with experiment design, theoretical validation of initial concepts, and contributed to the overall rigor of the paper. All authors contributed to interpreting results, editing, and revising the document.

\section*{Competing Interests}

The authors declare no competing interests.

\section*{Additional Information}

\noindent \textbf{Correspondence} and requests for materials should be addressed to Ethan N. Evans. Code will be made available upon request.

\end{document}

%% file: Packages.tex
\usepackage{setspace}

\usepackage[utf8]{inputenc} 
\usepackage[T1]{fontenc}    
\usepackage{url}            
\usepackage{booktabs}       
\usepackage{amsfonts}       
\usepackage{nicefrac}       
\usepackage{microtype}      
\usepackage{amsthm}
\usepackage{caption}
\usepackage{subcaption}
\usepackage{graphics} 
\usepackage{epsfig} 
\usepackage{times} 
\usepackage{amsmath} 
\usepackage{amssymb}  
\usepackage{makeidx}
\usepackage{float}
\usepackage{balance}
\usepackage{booktabs}
\usepackage{bbm}
\usepackage{mathrsfs}
\usepackage{enumerate}
\usepackage{multicol}
\usepackage{algorithm}
\usepackage{algpseudocode}

\usepackage{mathrsfs}  
\usepackage[normalem]{ulem}
\usepackage{xr} 
\usepackage{multicol}
\usepackage{float}
\usepackage[nolist]{acronym}
\usepackage{xcolor}
\usepackage[colorlinks]{hyperref}       

\hypersetup{
  colorlinks = true,
  allcolors = siaminlinkcolor,
  urlcolor = siamexlinkcolor,
}
\colorlet{siamexlinkcolor}{red!50!black} 
\colorlet{siaminlinkcolor}{black} 

\usepackage{lipsum} 

\usepackage{cleveref} 

\usepackage[acronym]{glossaries} 

\usepackage{titlesec} 

\usepackage{tabu}

\usepackage[compat=0.4]{yquant}
\usetikzlibrary{quotes}

\usepackage{scalerel}

%% file: Commands.tex
\newcommand{\rd}{{\mathrm d}}

\newcommand{\calD}{{\cal D}}
\newcommand{\calE}{{\cal E}}
\newcommand{\calF}{{\cal F}}

\newcommand{\calH}{{\cal H}}

\newcommand{\calK}{{\cal K}}

\newcommand{\calO}{{\cal O}}

\newcommand{\calU}{{\cal U}}

\newcommand{\T}{\mathsf{T}}

\newcommand{\Cb}{\mathbb{C}}

\newcommand{\Rb}{\mathbb{R}}

\newcommand{\tr}{{\mathrm {Tr}}}

\makeatletter
\renewcommand{\maketag@@@}[1]{\hbox{\m@th\normalsize\normalfont#1}}%
\makeatother

\catcode`\_=11\relax
    
    \def\_email#1@#2\q_nil{%
      \href{mailto:#1@#2}{{\emailfont #1\emailampersat #2}}
    }
    \newcommand\emailfont{}
    \newcommand\emailampersat{\small@}
    \catcode`\_=8\relax 

\newcommand{\parag}{\;\;\;\;}

\makeatletter
\newcommand{\vast}{\bBigg@{3.5}}
\newcommand{\Vast}{\bBigg@{4}}
\newcommand{\vastt}{\bBigg@{4.5}}
\newcommand{\Vastt}{\bBigg@{5}}
\makeatother

\newcommand{\ket}[1]{| #1 \rangle}
\newcommand{\bra}[1]{\langle #1 |}

\newcommand{\sprime}{\scaleobj{0.8}{\prime}}

\begin{acronym}
\acro{DNN}{Deep Neural Network}
\acro{ODE}{Ordinary Differential Equation}
\acro{SPDE}{Stochastic Partial Differential Equation}
\acro{FNN}{Feed-forward Neural Network}
\acro{CNN}{Convolutional Neural Network}
\acro{DP}{Dynamic Programming}
\acro{LSTM}{Long-Short Term Memory}
\acro{FC}{Fully Connected}
\acro{DDP}{Differential Dynamic Programming}
\acro{HJB}{Hamilton-Jacobi-Bellman}
\acro{PDE}{Partial Differential Equation}
\acro{PI}{Path Integral}
\acro{NN}{Neural Network}
\acro{SOC}{Stochastic Optimal Control}
\acro{RL}{Reinforcement Learning}
\acro{MPC}{Model Predictive Control}
\acro{IL}{Imitation Learning}
\acro{RNN}{Recurrent Neural Network}
\acro{DL}{Deep Learning}
\acro{SGD}{Stochastic Gradient Descent}
\acro{SDE}{Stochastic Differential Equation}
\acro{VRL}{Variational Reinforcement Learning}
\acro{IDVRL}{Infinite Dimensional Variational Reinforcement Learning}
\acro{1D}{1-dimensional}
\acro{2D}{2-dimensional}
\acro{ROM}{Reduced Order Model}
\acro{SDE}{Stochastic \ac{ODE}}
\end{acronym}

%% file: Acronyms.tex
\newacronym{ANN}{ANN}{Artificial Neural Network}
\newacronym{AIML}{AIML}{Artificial Intelligence and Machine Learning}
\newacronym{DNN}{DNN}{Deep Neural Network}
\newacronym{ODE}{ODE}{Ordinary Differential Equation}
\newacronym{SPDE}{SPDE}{Stochastic Partial Differential Equation}
\newacronym{FNN}{FNN}{Feed-forward Neural Network}
\newacronym{CNN}{CNN}{Convolutional Neural Network}
\newacronym{DP}{DP}{Dynamic Programming}
\newacronym{LSTM}{LSTM}{Long-Short Term Memory}
\newacronym{FC}{FC}{Fully Connected}
\newacronym{DDP}{DDP}{Differential Dynamic Programming}
\newacronym{HJB}{HJB}{Hamilton-Jacobi-Bellman}
\newacronym{PDE}{PDE}{Partial Differential Equation}
\newacronym{PI}{PI}{Path Integral}
\newacronym{NN}{NN}{Neural Network}
\newacronym{SOC}{SOC}{Stochastic Optimal Control}
\newacronym{RL}{RL}{Reinforcement Learning}
\newacronym{MPC}{MPC}{Model Predictive Control}
\newacronym{IL}{IL}{Imitation Learning}
\newacronym{RNN}{RNN}{Recurrent Neural Network}
\newacronym{DL}{DL}{Deep Learning}
\newacronym{SGD}{SGD}{Stochastic Gradient Descent}
\newacronym{SDE}{SDE}{Stochastic Differential Equation}
\newacronym{VRL}{VRL}{Variational Reinforcement Learning}
\newacronym{IDVRL}{IDVRL}{Infinite Dimensional Variational Reinforcement Learning}
\newacronym{ROM}{ROM}{Reduced Order Model}
\newacronym{ADPL}{ADPL}{Actuator Design and Policy Learning}
\newacronym{FBSDE}{FBSDE}{Forward-Backward Stochastic Differential Equation}
\newacronym{KL}{KL}{Kullback-Leibler}
\newacronym{DOF}{DOF}{Degrees of Freedom}
\newacronym{GRAPE}{GRAPE}{Gradient Ascent Pulse Engineering}
\newacronym{QND}{QND}{Quantum Non-Demolition}
\newacronym{MHD}{MHD}{magnetohydrodynamics}
\newacronym{STSO}{STSO}{Spatio-Temporal Stochastic Optimization}
\newacronym{LQR}{LQR}{Linear Quadratic Regulator}
\newacronym{STDDP}{STDDP}{Spatio-Temporal DDP}
\newacronym{RDE}{RDE}{Riccati Differential Equation}
\newacronym{1D}{1D}{1-dimensional}
\newacronym{2D}{2D}{2-dimensional}
\newacronym{3D}{3D}{3-dimensional}
\newacronym{RN}{RN}{Radon-Nikodym}
\newacronym{SNN}{SNN}{Sparse Neural Network}
\newacronym{FLOP}{FLOP}{Floating Point Operation}
\newacronym{GASS}{GASS}{Gradient-based Adaptive Stochastic Search}
\newacronym{NODE}{NODE}{Neural Ordinary Differential Equation}
\newacronym{QGASS}{QGASS}{Quantum Gradient-based Adaptive Stochastic Search}
\newacronym{MPPI}{MPPI}{Model Predictive Path Integral}
\newacronym{QVO-SS}{QVO-SS}{Quantum Variational Optimization - Single Shot}
\newacronym{QVO-MPC}{QVO-MPC}{Quantum Variational Optimization - MPC}
\newacronym{QSTSO}{QSTSO}{Quantum Spatio-Temporal Stochastic Optimization}
\newacronym{SME}{SME}{Stochastic Master Equation}
\newacronym{QRL}{QRL}{Quantum Reinforcement Learning}
\newacronym{ITC}{ITC}{Information Theoretic Control}
\newacronym{CFD}{CFD}{Computational Fluid Dynamics}
\newacronym{NP}{NP}{Non-Polynomial-Time}
\newacronym{ReLU}{ReLU}{Rectified Linear Unit}
\newacronym{NCOM}{NCOM}{Navy Coastal Ocean Model}
\newacronym{UUV}{UUV}{Unmanned Underwater Vehicle}
\newacronym{USV}{USV}{Unmanned Surface Vehicle}
\newacronym{LLM}{LLM}{Large Language Model}
\newacronym{QFT}{QFT}{quantum Fourier transform}
\newacronym{NISQ}{NISQ}{noisy intermediate scale quantum}
\newacronym{SASQuaTCh}{SASQuaTCh}{Self Attention Sequential Quantum Transformer Channel}
\newacronym{QSANN}{QSANN}{Quantum Self-Attention Neural Network}
\newacronym{GPT}{GPT}{Generative Pre-trained Transformer}
\newacronym{FNO}{FNO}{Fourier Neural Operator}
\newacronym{AFNO}{AFNO}{Adaptive Fourier Neural Operator}
\newacronym{DFT}{DFT}{Discrete Fourier Transform}
\newacronym{VQC}{VQC}{variational quantum circuit}
\newacronym{QML}{QML}{quantum machine learning}
\newacronym{SPSA}{SPSA}{Simultaneous Perturbation Stochastic Approximation}
\newacronym{MLP}{MLP}{Multi-Layer Perceptron}
\newacronym{RUS}{RUS}{repeat until success}
\newacronym{POVM}{POVM}{positive operator-valued measure}

\makeglossaries

%% file: QFVT.tex
\begin{array}{c}
\begin{tikzpicture}
\begin{yquant}
qubit {$r$} r[1];
qubit {$q_1$} a[1];
qubit {$\vdots$} b[1];
qubit {$q_s$} c[1];
qubit {$\vdots$} d[1];
qubit {$q_N$} e[1];
discard b;
discard d;

slash a;
slash c;
slash e;
hspace {5mm} -;

box {$\calE(x_1)$} a;
text {$\vdots$} b;
box {$\calE(x_s)$} c;
text {$\vdots$} d;
box {$\calE(x_N)$} e;
hspace {5mm} -;

box {$\text{QFT}^{\otimes N}$} (a,b,c,d,e);
hspace {5mm} -;

box {$U_{kernel}(\theta)$} (a,b,c,d,e);
hspace {5mm} -;

box {$\text{QFT}^{\dagger^{\otimes N}}$} (a,b,c,d,e);
h r;
hspace {5mm} -;

box {$U_{p}(\theta)$} r | a,c,e; 
hspace {5mm} -;
measure r;
hspace {5mm} -;
discard a,b,c,d,e;

\end{yquant}
\end{tikzpicture}
\end{array}

%% file: entangling_unitary.tex
\begin{array}{c}
\begin{tikzpicture}
\begin{yquant}
qubit {$q_1$} a[1];
qubit {$q_2$} b[1];
qubit {$q_3$} c[1];
qubit {$q_4$} d[1];
hspace {3mm} -;

[this subcircuit box style={dashed, "Strongly Entangling Layer 1"}]
subcircuit {
    [inout] 
    qubit {} a[1];
    qubit {} b[1];
    qubit {} c[1];
    qubit {} d[1];

    box {$R(\alpha_1^1, \beta_1^1, \gamma_1^1)$} a;
    box {$R(\alpha_2^1, \beta_2^1, \gamma_2^1)$} b;
    box {$R(\alpha_3^1, \beta_3^1, \gamma_3^1)$} c;
    box {$R(\alpha_4^1, \beta_4^1, \gamma_4^1)$} d;
    
    cnot b | a;
    hspace {0.5mm} -;
    cnot c | b;
    hspace {0.5mm} -;
    cnot d | c;
    hspace {0.5mm} -;
    cnot a | d;
    hspace {0.5mm} -;
} (a,b,c,d);

hspace {5mm} -;
[this subcircuit box style={dashed, "Strongly Entangling Layer 2"}]
subcircuit {
    [inout] 
    qubit {} a[1];
    qubit {} b[1];
    qubit {} c[1];
    qubit {} d[1];

    box {$R(\alpha_1^2, \beta_1^2, \gamma_1^2)$} a;
    box {$R(\alpha_2^2, \beta_2^2, \gamma_2^2)$} b;
    box {$R(\alpha_3^2, \beta_3^2, \gamma_3^2)$} c;
    box {$R(\alpha_4^2, \beta_4^2, \gamma_4^2)$} d;
    
    cnot b | a;
    hspace {0.5mm} -;
    cnot c | b;
    hspace {0.5mm} -;
    cnot d | c;
    hspace {0.5mm} -;
    cnot a | d;
    hspace {0.5mm} -;
} (a,b,c,d);
hspace {4mm} -;



\end{yquant}
\end{tikzpicture}
\end{array}

%% file: perceptron.tex
\begin{array}{c}
\begin{tikzpicture}
\begin{yquant}
qubit {$r$} r[1];
qubit {$q_1$} a[1];
qubit {$q_2$} b[1];
qubit {$\vdots$} c[1];
qubit {$q_N$} d[1];
discard c;

hspace {3mm} -;
box {$R_X$\\$\theta_1$} r | a;
hspace {1mm} -;
box {$R_X$\\ $\theta_2$} r;
hspace {1mm} -;
box {$R_Z$\\ $\theta_3$} r | a;
hspace {1mm} -;
box {$R_Z$\\$\theta_4$} r;
hspace {1mm} -;

box {$R_X$\\$\theta_5$} r | b;
hspace {1mm} -;
box {$R_X$\\ $\theta_6$} r;
hspace {1mm} -;
box {$R_Z$\\ $\theta_7$} r | b;
hspace {1mm} -;
box {$R_Z$\\$\theta_8$} r;
hspace {1mm} -;

text {$\,\,. \,\, .\,\, .\,\,$} r;
text {$\quad\vdots$} c;
hspace {1mm} -;

box {$R_X$\\$\theta_{4N-3}$} r | d;
hspace {1mm} -;
box {$R_X$\\ $\theta_{4N-2}$} r;
hspace {1mm} -;
box {$R_Z$\\ $\theta_{4N-1}$} r | d;
hspace {1mm} -;
box {$R_Z$\\$\theta_{4N}$} r;
hspace {3mm} -;

\end{yquant}
\end{tikzpicture}
\end{array}

%% file: Arxiv_Main_v2.bbl
\begin{thebibliography}{10}
\providecommand{\url}[1]{#1}
\csname url@samestyle\endcsname
\providecommand{\newblock}{\relax}
\providecommand{\bibinfo}[2]{#2}
\providecommand{\BIBentrySTDinterwordspacing}{\spaceskip=0pt\relax}
\providecommand{\BIBentryALTinterwordstretchfactor}{4}
\providecommand{\BIBentryALTinterwordspacing}{\spaceskip=\fontdimen2\font plus
\BIBentryALTinterwordstretchfactor\fontdimen3\font minus
  \fontdimen4\font\relax}
\providecommand{\BIBforeignlanguage}[2]{{%
\expandafter\ifx\csname l@#1\endcsname\relax
\typeout{** WARNING: IEEEtran.bst: No hyphenation pattern has been}%
\typeout{** loaded for the language `#1'. Using the pattern for}%
\typeout{** the default language instead.}%
\else
\language=\csname l@#1\endcsname
\fi
#2}}
\providecommand{\BIBdecl}{\relax}
\BIBdecl

\bibitem{vaswani2017attention}
A.~Vaswani, N.~Shazeer, N.~Parmar, J.~Uszkoreit, L.~Jones, A.~N. Gomez,
  {\L}.~Kaiser, and I.~Polosukhin, ``Attention is all you need,''
  \emph{Advances in neural information processing systems}, vol.~30, 2017.

\bibitem{achiam2023gpt}
J.~Achiam, S.~Adler, S.~Agarwal, L.~Ahmad, I.~Akkaya, F.~L. Aleman, D.~Almeida,
  J.~Altenschmidt, S.~Altman, S.~Anadkat \emph{et~al.}, ``{Gpt}-4 technical
  report,'' \emph{arXiv preprint arXiv:2303.08774}, 2023.

\bibitem{jiang2023mistral}
A.~Q. Jiang, A.~Sablayrolles, A.~Mensch, C.~Bamford, D.~S. Chaplot, D.~d.~l.
  Casas, F.~Bressand, G.~Lengyel, G.~Lample, L.~Saulnier \emph{et~al.},
  ``Mistral 7b,'' \emph{arXiv preprint arXiv:2310.06825}, 2023.

\bibitem{taori2023stanford}
R.~Taori, I.~Gulrajani, T.~Zhang, Y.~Dubois, X.~Li, C.~Guestrin, P.~Liang, and
  T.~B. Hashimoto, ``Stanford alpaca: an instruction-following llama model
  (2023),'' \emph{URL https://github. com/tatsu-lab/stanford\_alpaca}, vol.~1,
  no.~9, 2023.

\bibitem{kaplan2020scaling}
J.~Kaplan, S.~McCandlish, T.~Henighan, T.~B. Brown, B.~Chess, R.~Child,
  S.~Gray, A.~Radford, J.~Wu, and D.~Amodei, ``Scaling laws for neural language
  models,'' \emph{arXiv preprint arXiv:2001.08361}, 2020.

\bibitem{arora2023theory}
S.~Arora and A.~Goyal, ``A theory for emergence of complex skills in language
  models,'' \emph{arXiv preprint arXiv:2307.15936}, 2023.

\bibitem{dosovitskiy2021image}
A.~Dosovitskiy, L.~Beyer, A.~Kolesnikov, D.~Weissenborn, X.~Zhai,
  T.~Unterthiner, M.~Dehghani, M.~Minderer, G.~Heigold, S.~Gelly, J.~Uszkoreit,
  and N.~Houlsby, ``An image is worth 16x16 words: Transformers for image
  recognition at scale,'' in \emph{International Conference on Learning
  Representations}, 2021.

\bibitem{radford2023robust}
A.~Radford, J.~W. Kim, T.~Xu, G.~Brockman, C.~McLeavey, and I.~Sutskever,
  ``Robust speech recognition via large-scale weak supervision,'' in
  \emph{Proceedings of the 40th International Conference on Machine Learning},
  ser. ICML'23.\hskip 1em plus 0.5em minus 0.4em\relax JMLR.org, 2023.

\bibitem{geneva2022transformers}
N.~Geneva and N.~Zabaras, ``Transformers for modeling physical systems,''
  \emph{Neural Networks}, vol. 146, pp. 272--289, 2022.

\bibitem{shor1997polynomial}
P.~W. Shor, ``Polynomial-time algorithms for prime factorization and discrete
  logarithms on a quantum computer,'' \emph{SIAM Journal on Computing},
  vol.~26, no.~5, p. 1484–1509, Oct. 1997.

\bibitem{grover1996fast}
\BIBentryALTinterwordspacing
L.~K. Grover, ``A fast quantum mechanical algorithm for database search,'' in
  \emph{Proceedings of the Twenty-Eighth Annual ACM Symposium on Theory of
  Computing}, ser. STOC '96.\hskip 1em plus 0.5em minus 0.4em\relax New York,
  NY, USA: Association for Computing Machinery, 1996, p. 212–219. [Online].
  Available: \url{https://doi.org/10.1145/237814.237866}
\BIBentrySTDinterwordspacing

\bibitem{bennett2014cryptography}
C.~H. Bennett and G.~Brassard, ``Quantum cryptography: Public key distribution
  and coin tossing,'' \emph{Theoretical Computer Science}, vol. 560, pp. 7--11,
  2014, theoretical Aspects of Quantum Cryptography – celebrating 30 years of
  BB84.

\bibitem{preskill2018quantum}
J.~Preskill, ``Quantum {C}omputing in the {NISQ} era and beyond,''
  \emph{{Quantum}}, vol.~2, p.~79, Aug. 2018.

\bibitem{schuld2021machine}
M.~Schuld and F.~Petruccione, \emph{Machine Learning with Quantum Computers},
  ser. Quantum Science and Technology.\hskip 1em plus 0.5em minus 0.4em\relax
  Springer International Publishing, 2021.

\bibitem{goto2021universal}
T.~Goto, Q.~H. Tran, and K.~Nakajima, ``Universal approximation property of
  quantum machine learning models in quantum-enhanced feature spaces,''
  \emph{Physical Review Letters}, vol. 127, no.~9, p. 090506, 2021.

\bibitem{jager2023universal}
J.~J{\"a}ger and R.~V. Krems, ``Universal expressiveness of variational quantum
  classifiers and quantum kernels for support vector machines,'' \emph{Nature
  Communications}, vol.~14, no.~1, p. 576, 2023.

\bibitem{cerezo2022challenges}
M.~Cerezo, G.~Verdon, H.-Y. Huang, L.~Cincio, and P.~J. Coles, ``Challenges and
  opportunities in quantum machine learning,'' \emph{Nature Computational
  Science}, vol.~2, no.~9, pp. 567--576, 2022.

\bibitem{thanasilp2023subtleties}
S.~Thanasilp, S.~Wang, N.~A. Nghiem, P.~Coles, and M.~Cerezo, ``Subtleties in
  the trainability of quantum machine learning models,'' \emph{Quantum Machine
  Intelligence}, vol.~5, no.~1, p.~21, 2023.

\bibitem{liu2021rigorous}
Y.~Liu, S.~Arunachalam, and K.~Temme, ``A rigorous and robust quantum speed-up
  in supervised machine learning,'' \emph{Nature Physics}, vol.~17, no.~9, pp.
  1013--1017, 2021.

\bibitem{saggio2021experimental}
V.~Saggio, B.~E. Asenbeck, A.~Hamann, T.~Str{\"o}mberg, P.~Schiansky,
  V.~Dunjko, N.~Friis, N.~C. Harris, M.~Hochberg, D.~Englund \emph{et~al.},
  ``Experimental quantum speed-up in reinforcement learning agents,''
  \emph{Nature}, vol. 591, no. 7849, pp. 229--233, 2021.

\bibitem{lloyd2013quantum}
S.~Lloyd, M.~Mohseni, and P.~Rebentrost, ``Quantum algorithms for supervised
  and unsupervised machine learning,'' \emph{arXiv preprint arXiv:1307.0411},
  2013.

\bibitem{aimeur2013quantum}
E.~A{\"\i}meur, G.~Brassard, and S.~Gambs, ``Quantum speed-up for unsupervised
  learning,'' \emph{Machine Learning}, vol.~90, pp. 261--287, 2013.

\bibitem{li2020fourier}
Z.~Li, N.~Kovachki, K.~Azizzadenesheli, B.~Liu, K.~Bhattacharya, A.~Stuart, and
  A.~Anandkumar, ``Fourier neural operator for parametric partial differential
  equations,'' \emph{arXiv preprint arXiv:2010.08895}, 2020.

\bibitem{guibas2021adaptive}
J.~Guibas, M.~Mardani, Z.~Li, A.~Tao, A.~Anandkumar, and B.~Catanzaro,
  ``Adaptive {F}ourier neural operators: Efficient token mixers for
  transformers,'' \emph{arXiv preprint arXiv:2111.13587}, 2021.

\bibitem{pathak2022fourcastnet}
J.~Pathak, S.~Subramanian, P.~Harrington, S.~Raja, A.~Chattopadhyay,
  M.~Mardani, T.~Kurth, D.~Hall, Z.~Li, K.~Azizzadenesheli, P.~Hassanzadeh,
  K.~Kashinath, and A.~Anandkumar, ``{FourCastNet}: A global data-driven
  high-resolution weather model using adaptive {F}ourier neural operators,''
  \emph{arXiv preprint arXiv:2202.11214}, 2022.

\bibitem{leethorp2022fnet}
J.~Lee-Thorp, J.~Ainslie, I.~Eckstein, and S.~Ontanon, ``{FNet}: Mixing tokens
  with {F}ourier transforms,'' in \emph{Proceedings of the 2022 Conference of
  the North American Chapter of the Association for Computational Linguistics:
  Human Language Technologies}, 2022, pp. 4296--4313, arXiv preprint
  arXiv:2105.03824.

\bibitem{hales2000improved}
L.~Hales and S.~Hallgren, ``An improved quantum {F}ourier transform algorithm
  and applications,'' in \emph{Proceedings 41st Annual Symposium on Foundations
  of Computer Science}, 2000, pp. 515--525.

\bibitem{levin2023optimized}
\BIBentryALTinterwordspacing
M.~A. Levin, ``Optimized general uniform quantum state preparation,''
  \emph{Journal of Quantum Computing}, vol.~6, no.~1, p. 15–24, 2024.
  [Online]. Available: \url{http://dx.doi.org/10.32604/jqc.2024.047423}
\BIBentrySTDinterwordspacing

\bibitem{shukla2024}
A.~Shukla and P.~Vedula, ``An efficient quantum algorithm for preparation of
  uniform quantum superposition states,'' \emph{Quantum Information
  Processing}, vol.~23, no.~2, Jan. 2024.

\bibitem{li2022quantum}
G.~Li, X.~Zhao, and X.~Wang, ``Quantum self-attention neural networks for text
  classification,'' \emph{Science China Information Sciences}, vol.~67, p.
  142501, 2024, arXiv preprint arXiv:2205.05625.

\bibitem{zhao2024qksan}
R.-X. Zhao, J.~Shi, and X.~Li, ``Qksan: A quantum kernel self-attention
  network,'' \emph{IEEE Transactions on Pattern Analysis and Machine
  Intelligence}, 2024.

\bibitem{Liao2024GPT}
Y.~Liao and C.~Ferrie, ``{GPT on a Quantum Computer},'' \emph{arXiv preprint
  arXiv:2403.09418}, 3 2024.

\bibitem{xuemeasurement}
C.~Xue, Z.-Y. Chen, X.-F. Xu, X.-N. Zhuang, T.-P. Sun, Y.-J. Wang, J.~Wang,
  H.-Y. Liu, Y.-C. Wu, Z.~Wang \emph{et~al.}, ``Measurement information
  multiple-reuse allows deeper quantum transformer.''

\bibitem{schuld2021supervised}
M.~Schuld, ``Supervised quantum machine learning models are kernel methods,''
  \emph{arXiv preprint arXiv:2101.11020}, 2021.

\bibitem{laRose2020}
R.~LaRose and B.~Coyle, ``Robust data encodings for quantum classifiers,''
  \emph{Physical Review A}, vol. 102, no.~3, Sep. 2020.

\bibitem{lloyd2020quantum}
S.~Lloyd, M.~Schuld, A.~Ijaz, J.~Izaac, and N.~Killoran, ``Quantum embeddings
  for machine learning,'' \emph{arXiv preprint arXiv:2001.03622}, 2020.

\bibitem{amankwah2022quantum}
M.~G. Amankwah, D.~Camps, E.~W. Bethel, R.~Van~Beeumen, and T.~Perciano,
  ``Quantum pixel representations and compression for n-dimensional images,''
  \emph{Scientific reports}, vol.~12, no.~1, p. 7712, 2022.

\bibitem{mastriani2021quantum}
M.~Mastriani, ``Quantum {F}ourier transform is the building block for creating
  entanglement,'' \emph{Scientific Reports}, vol.~11, no.~1, p. 22210, 2021.

\bibitem{peruzzo2014variational}
A.~Peruzzo, J.~McClean, P.~Shadbolt, M.-H. Yung, X.-Q. Zhou, P.~J. Love,
  A.~Aspuru-Guzik, and J.~L. O’{B}rien, ``A variational eigenvalue solver on
  a photonic quantum processor,'' \emph{Nature communications}, vol.~5, no.~1,
  p. 4213, 2014.

\bibitem{mitarai2018}
K.~Mitarai, M.~Negoro, M.~Kitagawa, and K.~Fujii, ``Quantum circuit learning,''
  \emph{Physical Review A}, vol.~98, no.~3, Sep. 2018.

\bibitem{schuld2019}
M.~Schuld, V.~Bergholm, C.~Gogolin, J.~Izaac, and N.~Killoran, ``Evaluating
  analytic gradients on quantum hardware,'' \emph{Physical Review A}, vol.~99,
  no.~3, Mar. 2019.

\bibitem{banchi2021measuring}
L.~Banchi and G.~E. Crooks, ``Measuring analytic gradients of general quantum
  evolution with the stochastic parameter shift rule,'' \emph{Quantum}, vol.~5,
  p. 386, 2021.

\bibitem{spall1998overview}
J.~C. Spall, ``An overview of the simultaneous perturbation method for
  efficient optimization,'' \emph{Johns Hopkins apl technical digest}, vol.~19,
  no.~4, pp. 482--492, 1998.

\bibitem{gacon2021simultaneous}
J.~Gacon, C.~Zoufal, G.~Carleo, and S.~Woerner, ``Simultaneous perturbation
  stochastic approximation of the quantum {F}isher information,''
  \emph{Quantum}, vol.~5, p. 567, 2021.

\bibitem{spall2000adaptive}
J.~C. Spall, ``Adaptive stochastic approximation by the simultaneous
  perturbation method,'' \emph{IEEE transactions on automatic control},
  vol.~45, no.~10, pp. 1839--1853, 2000.

\bibitem{berahas2016multi}
A.~S. Berahas, J.~Nocedal, and M.~Tak{\'a}c, ``A multi-batch {L-BFGS} method
  for machine learning,'' \emph{Advances in Neural Information Processing
  Systems}, vol.~29, 2016.

\bibitem{kingma2014adam}
D.~P. Kingma and J.~Ba, ``Adam: A method for stochastic optimization,''
  \emph{arXiv preprint arXiv:1412.6980}, 2014.

\bibitem{bergholm2018pennylane}
V.~Bergholm, J.~Izaac, M.~Schuld, C.~Gogolin, S.~Ahmed, V.~Ajith, M.~S. Alam,
  G.~Alonso-Linaje, B.~AkashNarayanan, A.~Asadi, J.~M. Arrazola, U.~Azad,
  S.~Banning, C.~Blank, T.~R. Bromley, B.~A. Cordier, J.~Ceroni, A.~Delgado,
  O.~D. Matteo, A.~Dusko, T.~Garg, D.~Guala, A.~Hayes, R.~Hill, A.~Ijaz,
  T.~Isacsson, D.~Ittah, S.~Jahangiri, P.~Jain, E.~Jiang, A.~Khandelwal,
  K.~Kottmann, R.~A. Lang, C.~Lee, T.~Loke, A.~Lowe, K.~McKiernan, J.~J. Meyer,
  J.~A. Montañez-Barrera, R.~Moyard, Z.~Niu, L.~J. O'Riordan, S.~Oud,
  A.~Panigrahi, C.-Y. Park, D.~Polatajko, N.~Quesada, C.~Roberts, N.~Sá,
  I.~Schoch, B.~Shi, S.~Shu, S.~Sim, A.~Singh, I.~Strandberg, J.~Soni,
  A.~Száva, S.~Thabet, R.~A. Vargas-Hernández, T.~Vincent, N.~Vitucci,
  M.~Weber, D.~Wierichs, R.~Wiersema, M.~Willmann, V.~Wong, S.~Zhang, and
  N.~Killoran, ``{PennyLane}: Automatic differentiation of hybrid
  quantum-classical computations,'' \emph{arXiv preprint arXiv:1811.04968},
  2018.

\bibitem{schuld2020circuit}
M.~Schuld, A.~Bocharov, K.~M. Svore, and N.~Wiebe, ``Circuit-centric quantum
  classifiers,'' \emph{Physical Review A}, vol. 101, no.~3, Mar. 2020.

\bibitem{maronese2022quantum}
M.~Maronese, C.~Destri, and E.~Prati, ``Quantum activation functions for
  quantum neural networks,'' \emph{Quantum Information Processing}, vol.~21,
  no.~4, p. 128, 2022.

\bibitem{gili2023introducing}
K.~Gili, M.~Sveistrys, and C.~Ballance, ``Introducing nonlinear activations
  into quantum generative models,'' \emph{Physical Review A}, vol. 107, no.~1,
  p. 012406, 2023.

\bibitem{holmes2023nonlinear}
Z.~Holmes, N.~J. Coble, A.~T. Sornborger, and Y.~Suba{\c{s}}{\i}, ``Nonlinear
  transformations in quantum computation,'' \emph{Physical Review Research},
  vol.~5, no.~1, p. 013105, 2023.

\bibitem{lecun2010mnist}
Y.~LeCun, C.~Cortes, and C.~Burges, ``{MNIST} handwritten digit database,''
  \emph{ATT Labs [Online]. Available: http://yann.lecun.com/exdb/mnist},
  vol.~2, 2010.

\bibitem{Paszke2019PyTorch}
A.~Paszke, S.~Gross, F.~Massa, A.~Lerer, J.~Bradbury, G.~Chanan, T.~Killeen,
  Z.~Lin, N.~Gimelshein, L.~Antiga, A.~Desmaison, A.~Köpf, E.~Yang, Z.~DeVito,
  M.~Raison, A.~Tejani, S.~Chilamkurthy, B.~Steiner, L.~Fang, J.~Bai, and
  S.~Chintala, ``Pytorch: An imperative style, high-performance deep learning
  library,'' \emph{Advances in neural information processing systems}, vol.~32,
  pp. 8026--8037, 2019.

\bibitem{bermejo2024quantum}
P.~Bermejo, P.~Braccia, M.~S. Rudolph, Z.~Holmes, L.~Cincio, and M.~Cerezo,
  ``Quantum convolutional neural networks are (effectively) classically
  simulable,'' \emph{arXiv preprint arXiv:2408.12739}, 2024.

\bibitem{Our_comment}
The PennyLane package used in our implementation is limited by a Numpy
  implementation that is constrained to 32 dimensions. This is because the
  PennyLane implementation uses one dimension of the Numpy array per qubit. To
  accurately embed and test on the CIFAR dataset, one would need more than 32
  qubits, and thus a Numpy array of greater than 32 dimensions. A potential
  work-around is to switch to another quantum programming library, such as
  Qiskit or Q\#, while simultaneously retaining the interface with PyTorch.

\bibitem{plesch2011quantum}
M.~Plesch and {\v{C}}.~Brukner, ``Quantum-state preparation with universal gate
  decompositions,'' \emph{Physical Review A—Atomic, Molecular, and Optical
  Physics}, vol.~83, no.~3, p. 032302, 2011.

\bibitem{nakaji2022approximate}
K.~Nakaji, S.~Uno, Y.~Suzuki, R.~Raymond, T.~Onodera, T.~Tanaka, H.~Tezuka,
  N.~Mitsuda, and N.~Yamamoto, ``Approximate amplitude encoding in shallow
  parameterized quantum circuits and its application to financial market
  indicators,'' \emph{Physical Review Research}, vol.~4, no.~2, p. 023136,
  2022.

\bibitem{salvi2022neural}
C.~Salvi, M.~Lemercier, and A.~Gerasimovics, ``Neural stochastic pdes:
  Resolution-invariant learning of continuous spatiotemporal dynamics,''
  \emph{Advances in Neural Information Processing Systems}, vol.~35, pp.
  1333--1344, 2022.

\end{thebibliography}
